\documentclass{aa}  

\usepackage{graphicx}
\usepackage[varg]{txfonts}
\usepackage{color}

\begin{document}

   \title{A low-luminosity type-1 QSO sample}

   \subtitle{
   II. Tracing circumnuclear star formation in HE 1029-1831 with SINFONI
   \thanks{Based on observations with ESO-VLT, proposal no. 093.B-0718(A)}}

   \author{Gerold Busch\inst{1}, Semir Smaji\'c\inst{1,2}, Julia Scharw\"{a}chter\inst{3}, Andreas Eckart\inst{1,2}, M\'{o}nica Valencia-S.\inst{1}, Lydia Moser\inst{1}, Bernd Husemann\inst{4}, Melanie Krips\inst{5}, \and Jens Zuther\inst{1}
          }

   \institute{I. Physikalisches Institut, Universit\"at zu K\"oln,
              Z\"ulpicher Str. 77, 50937 K\"oln, Germany \\
              \email{busch@ph1.uni-koeln.de}
              \and
              Max-Planck-Institut f\"ur Radioastronomie,
              Auf dem H\"ugel 69, 53121 Bonn, Germany
              \and
              LERMA (CNRS: UMR 8112), Observatoire de Paris, 61 Av. de l'Observatoire, 75014 Paris, France
              \and
              European Southern Observatory, Karl-Schwarzschild-Str. 2, 85748 Garching b. München, Germany
              \and
              Institut de Radio Astronomie Millimétrique (IRAM), 300 rue de la Piscine, Domaine Universitaire de Grenoble, 38406 St. Martin d'Hères, France
             }

   \date{Received; }

  \abstract{
Circumnuclear star formation and AGN feedback is believed to play a critical role in the context of galaxy evolution. 
The low-luminosity QSO (LLQSO) sample that contains 99 of the closest AGN with redshift $z\leq 0.06$ fills the gap between the local AGN population and high-redshift QSOs that is essential to understand the AGN evolution with redshift.
In this paper, we present the results of near-infrared $H$- and $K$- integral field spectroscopy of the inner kiloparsecs of the LLQSO HE 1029-1831 with SINFONI. Line maps show that ionized hydrogen gas is located in spiral arms within the stellar bar and a circumnuclear ring. Line fluxes and diagnostic line ratios indicate recent or ongoing star formation in the circumnuclear region and the presence of young and intermediate-age stellar populations in the bulge. In particular, we find traces of an intense starburst in the circumnuclear region that has begun around 100 Myr ago but has declined to a fraction of the maximum intensity now. We estimate the dynamical bulge mass and find that the galaxy follows published $M_\mathrm{BH}-M_\mathrm{bulge}$ relations. However, bulge-disk decomposition of the $K$-band image with \textsc{Budda} reveals that HE 1029-1831 does not follow the $M_\mathrm{BH}-L_\mathrm{bulge}$ relations of inactive galaxies. We conclude that the deviation from $M_\mathrm{BH}-L_\mathrm{bulge}$ relations of inactive galaxies in this source is rather caused by young stellar populations and not by an undermassive black hole.
}


\titlerunning{LLQSO sample - Tracing circumnuclear star formation in HE 1029-1831 with SINFONI}
\authorrunning{Gerold Busch et al.}

\maketitle

\section{Introduction}

Tight correlations between the mass of the central supermassive black hole (SMBH) and properties of the hosting galaxy, particularly its central bulge component, have been found \citep{1998AJ....115.2285M,2000ApJ...539L...9F,2003ApJ...589L..21M,2004ApJ...604L..89H,2007ApJ...655...77G,2013MNRAS.434..387S} and are seen as evidence for a co-evolution of black holes and their host galaxies. More recent studies have revealed galaxies that do not follow these relations: \cite{2011Natur.469..374K} discuss that only ``classical bulges'' that formed by galaxy mergers show correlations with the SMBH while disk-like ``pseudobulges'' that are forming by secular evolution do not. Other authors suggest that different formation processes (gaseous formation processes vs. dry merging) might become visible in differing SMBH-bulge scaling relations \citep{2013ApJ...764..151G,2013ApJ...768...76S}. Further, \cite{2014ApJ...780...70L} claim that, providing a thorough decomposition, the correlation between black hole mass and total host galaxy luminosity could be equally tight as the correlation between black hole mass and bulge luminosity.

In our recent study \citep{2014A&A...561A.140B}, we probed the $M_\mathrm{BH}-L_\mathrm{bulge}$ relation of low-luminosity type-1 QSOs (LLQSOs). We performed a careful decomposition of $K$-band images with the \textsc{Budda}-code \citep{2004ApJS..153..411D,2008MNRAS.384..420G} and find that the bulges do not follow the published $M_\mathrm{BH}-L_\mathrm{bulge}$ relations for inactive galaxies of \cite{2003ApJ...589L..21M,2012MNRAS.419.2264V,2013ApJ...764..151G,2013ARA&A..51..511K}. A deviation of type-1 AGN from the SMBH-bulge relations is in agreement with previous studies in the optical \citep{2004ApJ...615..652N,2008ApJ...687..767K,2011ApJ...726...59B}. Assuming that a SMBH-bulge relation traces a co-evolution, this could hint at a different evolution of inactive and active galaxies. On the other hand, a deviation from the relations of inactive galaxies could hint at different properties of active galaxies compared to inactive ones, caused by direct interaction through AGN feedback. Particularly, we discussed an undermassive black hole that is growing ``towards'' the relation or an overluminosity of the bulge produced by young and/or intermediate-age stellar populations. With the emergence of AO-assisted integral-field spectroscopy, many studies have been focused on the circumnuclear star formation properties in AGN \cite[e.g.,][]{2007A&A...466..451Z,2007ApJ...671.1388D,2008AJ....135..479B,2010MNRAS.404..166R}. However, the impact on the bulge and overall host galaxy and the interaction between central star formation and AGN feedback is still under discussion.

\subsection{The low-luminosity type-1 QSO sample}

The low-luminosity type-1 QSO sample (LLQSO sample) is a subsample of the Hamburg/ESO survey for bright UV-excess QSOs \citep{2000A&A...358...77W} that contains only the 99 brightest and nearest QSOs with a redshift $z\leq 0.06$. In many properties, e.g. redshift, gas masses, and luminosities, the LLQSO-sample lies between the NUGA sample \citep[NUclei of GAlaxies; e.g.,][]{2003A&A...407..485G,2004A&A...414..857C,2005A&A...441.1011G,2007A&A...464..553K} and the Palomar Green (PG) QSO sample. 
This connects the advantages of cosmological proximity and of higher spatial coverage. The first enables us to observe the processes in the centers of the galaxies in detail with high spatial resolution. The latter raises the probability of observing QSOs with high luminosity and accretion rate \citep{2007A&A...464..187K,2007A&A...470..571B,2012nsgq.confE..69M}. 

In Table \ref{tab:comp_qso}, we compare the source HE 1029-1831 with NGC 3227, the nearest Seyfert-1 galaxy, and 3C 273, the nearest quasar. We see that the LLQSO has a much higher bolometric luminosity and accretion rate (traced by the Eddington ratio) than the nearby Seyfert galaxy. However, it is much closer than the next quasar, the resolution being better by a factor of four. Thus, LLQSOs form an ideal sample to study active galaxies with significant accretion activity at a resolution that is still good enough for detailed analysis.

\cite{2007A&A...470..571B} and \cite{2007A&A...464..187K} observed 39 galaxies from the LLQSO sample in CO from which 27 have been detected. They report that most galaxies are rich in molecular gas, with a range of gas masses from $0.4\times 10^9\,M_\odot$ to $9.7\times 10^9\,M_\odot$. Furthermore, they find that the gas mass is correlated with the star formation and AGN activity as indicated by the far-infrared luminosity. \cite{2009A&A...507..757K} searched for 21 cm \ion{H}{i} emission in these 27 CO-detected galaxies with the Effelsberg 100m-telescope. They find neutral atomic gas masses ranging from $1.1\times 10^9\,M_\odot$ to $3.8\times 10^{10}\,M_\odot$ but no strong correlation between gas mass and infrared emission that traces star formation and AGN activity. \cite{2006A&A...452..827F} observed nine galaxies from the sample with the near-infrared spectrograph ISAAC (VLT) and find that the spectra show signs for both, AGN activity but also star formation. In our near-infrared study \citep{2012nsgq.confE..60B,2014A&A...561A.140B} we find that the LLQSOs do not follow published $M_\mathrm{BH}-L_\mathrm{bulge}$ relations for inactive galaxies. A reason could be massive star formation in the circumnuclear regions. Integral-field spectroscopy with SINFONI is the preferred tool to examine the excitation mechanisms of emission lines as well as the kinematics of gas and stars in a spatially resolved way and learn about the properties of star formation and AGN in the centers of LLQSOs. 

Since HE 1029-1831 shows strong CO(1-0) emission \citep[][corresponding to a cold gas mass of $\sim 1.2\times 10^{10}\,M_\odot$]{2007A&A...464..187K}, it is an ideal object for a pilot study to demonstrate that with SINFONI we can successfully trace star formation regions as well as hot molecular and ionized gas reservoirs also in LLQSOs which are more distant than most of the previously analyzed (e.g. by the NUGA team) objects.

\begin{table*}
\centering
\caption{Comparison of the nearest Seyfert-1 galaxy NGC 3227, our source HE 1029-1831, and the nearest quasar 3C 273.} 
\label{tab:comp_qso}
\begin{tabular}{ccccccc} \hline \hline
source & redshift & scale & resolution\tablefootmark{(*)} & $\log(L_\mathrm{bol}[\mathrm{W}])$  & $\log(M_\mathrm{BH}[M_\odot])$ & $\lambda_\mathrm{Edd}$ \\ \hline

NGC 3227 & 0.004 & $100\,\mathrm{pc}/\arcsec$ & $13\,\mathrm{pc}$ & 36.86 & 7.64 & 0.013 \\
HE 1029-1831 & 0.0404 & $800\,\mathrm{pc}/\arcsec$ & $110\,\mathrm{pc}$ & 37.6 & 7.2 & 0.3 \\
3C 273 & 0.158 & $3150\,\mathrm{pc}/\arcsec$ & $410\,\mathrm{pc}$ & 39.94 & 8.95 & 0.8 \\

\hline
\end{tabular}
\tablefoot{\tablefoottext{*}{Assuming a resolution of $0\farcs13$ as we reach with AO correction in our observations with the $3\arcsec \times 3\arcsec$ FOV. The redshift is taken from the NED database. Black hole mass and bolometric luminosity of NGC 3227 and 3C 273 are from \cite{2002ApJ...579..530W} and references therein.}}
\end{table*}

\subsection{HE1029-1831}

\begin{figure}
\includegraphics[width=\columnwidth]{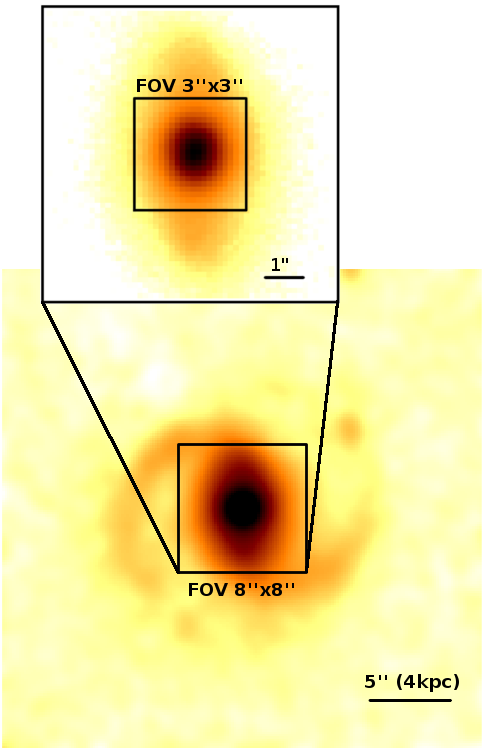}
\caption{ISAAC $K$-band image of HE1029-1831. The $8\arcsec \times 8\arcsec$ and $3\arcsec \times 3\arcsec$ FOVs of our SINFONI observations are marked. The image has been smoothed with a 3-pixel Gaussian and cleared from bad pixels.}
\label{fig:isaac}
\end{figure}

\cite{2006A&A...452..827F} analyze NIR $JHK$-images as well as $H+K$-longslit spectra of nine LLQSOs, including HE 1029-1831, obtained with ISAAC at the VLT. In Fig. \ref{fig:isaac}, we see that HE 1029-1831 is a barred spiral galaxy with two prominent spiral arms. The bar has a length of about $8\arcsec \approx 6.4\,\mathrm{kpc}$.

In their spectra, they can separate the Pa$\alpha$ line into broad and narrow component; the broad component has a full width at half maximum (FWHM) of $2081\,\mathrm{km}\,\mathrm{s}^{-1}$. Furthermore, they find extended molecular hydrogen emission. Due to the FWHM(Pa$\alpha$) of about $2000\,\mathrm{km}\,\mathrm{s}^{-1}$, they classify HE 1029-1831 as narrow-line Seyfert-1 galaxy (NLSy1). This is supported by \cite{2000ApJS..126...63R}, using the name ``CTS J04.08'', who find a FWHM(H$\alpha$) of $1870\,\mathrm{km}\,\mathrm{s}^{-1}$ \citep[but see discussion about NLSy1s in][]{2012nsgq.confE..17V}.

\cite{2007A&A...464..187K} observed HE 1029-1831 in the CO(1-0) and CO(2-1) line emission with the IRAM Plateau de Bure Interferometer and obtain a total gas mass of $M(\mathrm{H}_2+\mathrm{He})\approx 8 \times 10^9\,M_\odot$. For the spatial extent of the emission, they find a FWHM of $\approx(7\pm 2)\arcsec$, which means that the observed CO emission comes from a region including the complete stellar bar. The \ion{H}{i} mass has been measured from observations with the Effelsberg 100m-telescope and is $(6.6\pm0.6)\times 10^9\,M_\odot$ \citep{2009A&A...507..757K}.

From optical line ratios, HE 1029-1831 is classified as Seyfert 1.5 galaxy in the 12th catalog of quasars and AGN by \cite{2006A&A...455..773V}. \cite{2001ApJS..132...37K} took optical spectra and determine the position of the galaxy in optical diagnostic diagrams (using the name IRAS 10295-1831). Depending on the diagram, the galaxy is classified as AGN, starburst or composite. From this, we expect that in HE 1029-1831 contributions from both, AGN and starburst, are important.\\

In this paper, we present near-infrared integral-field spectroscopy of HE 1029-1831 with SINFONI. HE 1029-1831 is the first of a set of galaxies from the LLQSO sample that has been observed with SINFONI. From the 3D-spectroscopy, we aim to learn about the stellar and gas masses as well as kinematics in LLQSOs. We will determine reliable mass-to-light ratios and trace the star formation history in the centers of type-1 AGN. 

The paper is organized as follows: In Sect. \ref{sec:obs}, we present the observations with SINFONI as well as preceding observations with ISAAC (near-infrared image) and describe our reduction procedures. In Sect. \ref{sec:results}, we describe our methods, that is the bulge-disk decomposition with \textsc{Budda}, emission line fitting, and stellar continuum subtraction with pPXF, and present the results. In Sect. \ref{sec:discussion}, we discuss the results, particularly we estimate black hole masses, determine gas masses and gas excitation mechanisms, and discuss the impact of star formation. In Sect. \ref{sec:summary}, we give a short summary and conclusions.

Throughout this paper, we use a standard cosmology with $H_0=70\,\mathrm{km}\,\mathrm{s}^{-1}\,\mathrm{Mpc}^{-1}$, $\Omega_\mathrm{m}=0.3$, and $\Omega_\Lambda=0.7$. At a redshift of $z=0.0404$, this results in a luminosity distance of $D_L=177\,\mathrm{Mpc}$ and a scale of $0.8\,\mathrm{kpc}/\arcsec$.

\section{Observation and data reduction}
\label{sec:obs}

\subsection{SINFONI integral-field spectroscopy}
\label{sec:obs_sinfo}
We present results of high-resolution NIR observations of HE 1029-1831 that have been carried out April 20-22, 2014 with SINFONI at the Unit Telescope 4 of the ESO Very Large Telescope in Chile \citep[VLT,][]{2003SPIE.4841.1548E,2004Msngr.117...17B}.

We started the observations with the $0\farcs25$ plate scale, corresponding to a field-of-view (FOV) of $8\arcsec \times 8\arcsec$ without adaptive optics. Then, we used the $0\farcs1$ plate scale with adaptive optics assistance, resulting in higher spatial resolution but with a smaller FOV of $3\arcsec \times 3\arcsec$. We used the $H+K$ grating providing a spectral resolution of $R_{H+K}=1500$. A TST... (T: target, S: sky) pattern with 150s integration time each was used to produce sky-subtracted frames. Furthermore, a jitter pattern with offsets of $\pm 1 \arcsec$ for the $8\arcsec \times 8\arcsec$-FOV and $\pm 0\farcs5$ for the $3\arcsec \times 3\arcsec$-FOV was used to minimize the impact of dead pixels. The total integration time on source was $4500\,\mathrm{s}$ for the large and $3000\,\mathrm{s}$ for the small FOV, with additional $2250\,\mathrm{s}$ and $1500\,\mathrm{s}$ for sky observations. In between, telluric standards have been observed.

In the raw files, detector specific problems occured that we corrected following the procedure described in the appendix of \cite{2014A&A...567A.119S}. Then, we used the SINFONI pipeline for data reduction up to single-exposure-cube reconstruction. The alignment and final coaddition of the single-exposure-cubes has been performed with our own dpuser\footnote{http://www.mpe.mpg.de/\~{}ott/dpuser/dpuser.html} routines.

For the telluric correction, we used G2V stars that we have observed in between. A high-quality solar spectrum \citep{1996AJ....111..537M} that has been convolved down to the resolution of the telluric spectrum was used to correct the features of the G2V star. At the spectral edges and in the spectral region between $H$- and $K$-band, the solar spectrum had to be interpolated by a black body function with $T=5800\,\mathrm{K}$.

The FWHM of the point spread function was measured by fitting Gaussians to the telluric standard stars that are supposed to be point sources. We measure a FWHM of $0\farcs4 - 0\farcs6$ for the FOV of $8\arcsec \times 8\arcsec$, which corresponds to linear scales of $320 - 500$ pc. For the FOV of $3\arcsec \times 3\arcsec$, we measure a FWHM of $0\farcs13$, corresponding to $110$ pc. 

Calibration was done during the telluric correction. We summed up the counts of the telluric standard star within a radius of three times the FWHM of the PSF, centered on the peak position. The standard star counts at $\lambda 2.159\,\mu\mathrm{m}$ were taken as a reference for the science target. The magnitude of the standard star was taken from the 2MASS point source catalog \citep{2006AJ....131.1163S}. In order to convert the 2MASS magnitude into a flux density, we use the Spitzer Science Center Magnitude to Flux Density converter\footnote{http://ssc.spitzer.caltech.edu/warmmission/propkit/pet/magtojy/}.

\subsection{ISAAC $K$-band imaging}
A $K$-band image of HE1029-1831, observed with ISAAC \citep{1998Msngr..94....7M} at the VLT on April 20, 2003, is available in the ESO-archive. ISAAC provides a FOV of $152\arcsec \times 152\arcsec$ with a pixelscale of $0.1484\arcsec/\mathrm{pixel}$. Four frames have been observed in a TSST jitter pattern with an exposure time of $2\sec$ each. After subtracting the sky frames, the two frames have been shifted to a common reference frame and a mean image was created. Calibration was done using 2MASS magnitudes of foreground stars. For details on the data, see \cite{2006A&A...452..827F}.

\section{Results}
\label{sec:results}

\subsection{Bulge-Disk Decomposition}
\label{sec:decomp}
Parametric modeling of galaxy images is an important tool to disentangle the light of different galaxy components, such as bulge, disk, bar, AGN, to measure their structural parameters and determine light fractions of the galactic components. These values are essential for estimates of stellar masses and are used to establish relations e.g. between black hole mass and bulge luminosity.

In this work, we use \textsc{Budda}\footnote{http://www.sc.eso.org/\~{}dgadotti/budda.html} \citep[Bulge/Disk Decomposition Analysis;][]{2004ApJS..153..411D,2008MNRAS.384..420G} for the two-dimensional decomposition of the $K$-band image.

To model the disk component, we use an exponential function 
\begin{equation}
\mu_\mathrm{disk}(r) = \mu_0 + 1.086 \frac{r}{h_r}
\end{equation}
with central surface density $\mu_0$ and the characteristic scale-length $h_r$ \citep{1970ApJ...160..811F}.

Bulge and bar component are modeled by a S\'ersic function
\begin{equation}
\mu(r) = \mu_e + c_n \left[ \left( \frac{r}{r_e} \right)^{1/n} -1 \right]
\end{equation}
with an effective radius $r_e$ that is the radius that includes half of the light emitted by the respective component (``half-light radius'') and the effective surface brightness $\mu_e\equiv \mu(r_e)$. $n$ is the S\'ersic index and $c_n=2.5(0.868 n -0.142)$ is a parameter that depends on $n$ \citep{1968adga.book.....S,1993MNRAS.265.1013C}. 

The AGN component is modeled by a Moffat function with width fixed to that of the seeing PSF (measured by fitting Gaussian functions to point sources). The width is therefore kept fixed and only the peak intensity value is varied. \textsc{Budda} calculates the minimum valid count as the square root of the noise level. Here, the limiting surface brightness is $\mu_K=21.5\,\mathrm{mag}\,\mathrm{arcsec}^{-2}$. All pixels with count level below will be masked. \cite{2014A&A...561A.140B} give more details on the fitting process and limits of 2D-decomposition.

\begin{figure*}
\includegraphics[width=\linewidth]{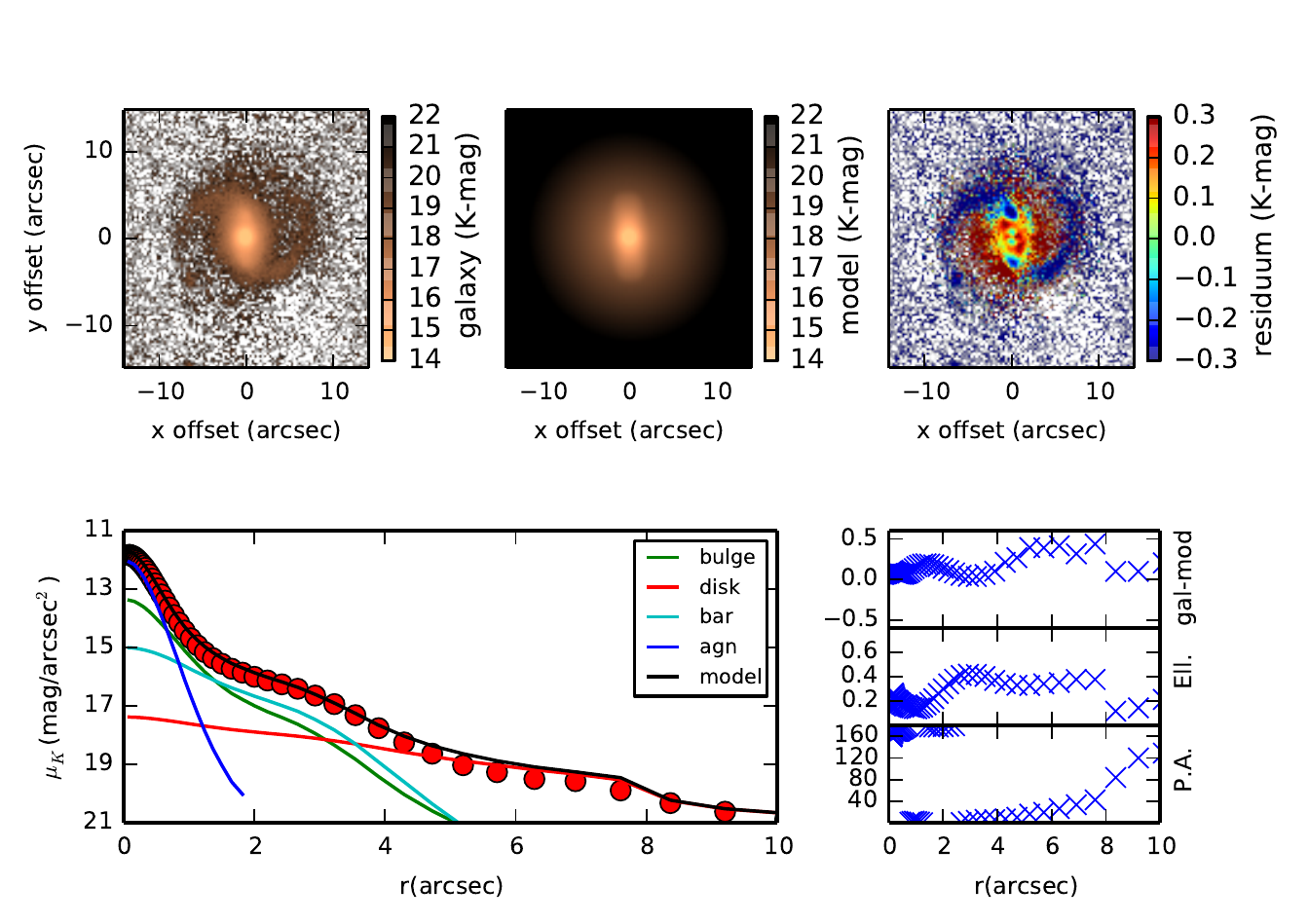}
\caption{Decomposition of HE 1029-1831 with \textsc{Budda}. We show (\emph{from left to right}) the original $K$-band ISAAC-image, the \textsc{Budda}-model and the residuum obtained by division galaxy/model. Here, blue indicates that the original image has more flux than the model while red indicates that the model has more flux than the original image. In the \emph{lower row}, we show an elliptically averaged radial profile of the galaxy, the single components and the model. Next to it, we show the difference galaxy-model, ellipticity and position angle.}
\label{fig:buddafit}
\end{figure*}

Figure \ref{fig:buddafit} shows the results of the 2D-decomposition of HE 1029-1831: In the upper row, from left to right the original $K$-band image, the \textsc{Budda} model image and a residual image. In the lower row, we show radial profiles that have been extracted using the \textsc{Ellipse} task of \textsc{Iraf}. In the fits of the model components, ellipticity and position angle have been fixed to the values of the original image.

The fit results in a bulge component with an effective radius of $r_e=0.59\arcsec$ ($=0.47\,\mathrm{kpc}$) and an apparent magnitude of $m_\mathrm{bulge}=12.69$ (absolute magnitude: $M_\mathrm{bulge}=-23.57$). The disk has a scale length of $h_r=2.70\arcsec$ ($=2.16\,\mathrm{kpc}$) and a magnitude of $m_\mathrm{disk}=13.04$ ($M_\mathrm{disk}=-23.22$). The luminosity fractions are: 29\% (bulge), 21\% (disk), 22\% (bar), and 28\% (AGN).

The S\'ersic index of the bulge is $n=1.2$. A S\'ersic index below 2 is indicative for a disk-like pseudobulge \citep{2004ARA&A..42..603K}. However, on the one hand the S\'ersic index should not be used as a sole indicator \citep{2009MNRAS.393.1531G} and on the other hand, since the central light distribution is heavily affected by the nuclear point source, it is almost impossible to determine the S\'ersic index reliably.

\subsection{Emission-line fits}
\label{sec:emline-fluxes}
\begin{figure*}
\centering
\includegraphics[width=0.45\linewidth]{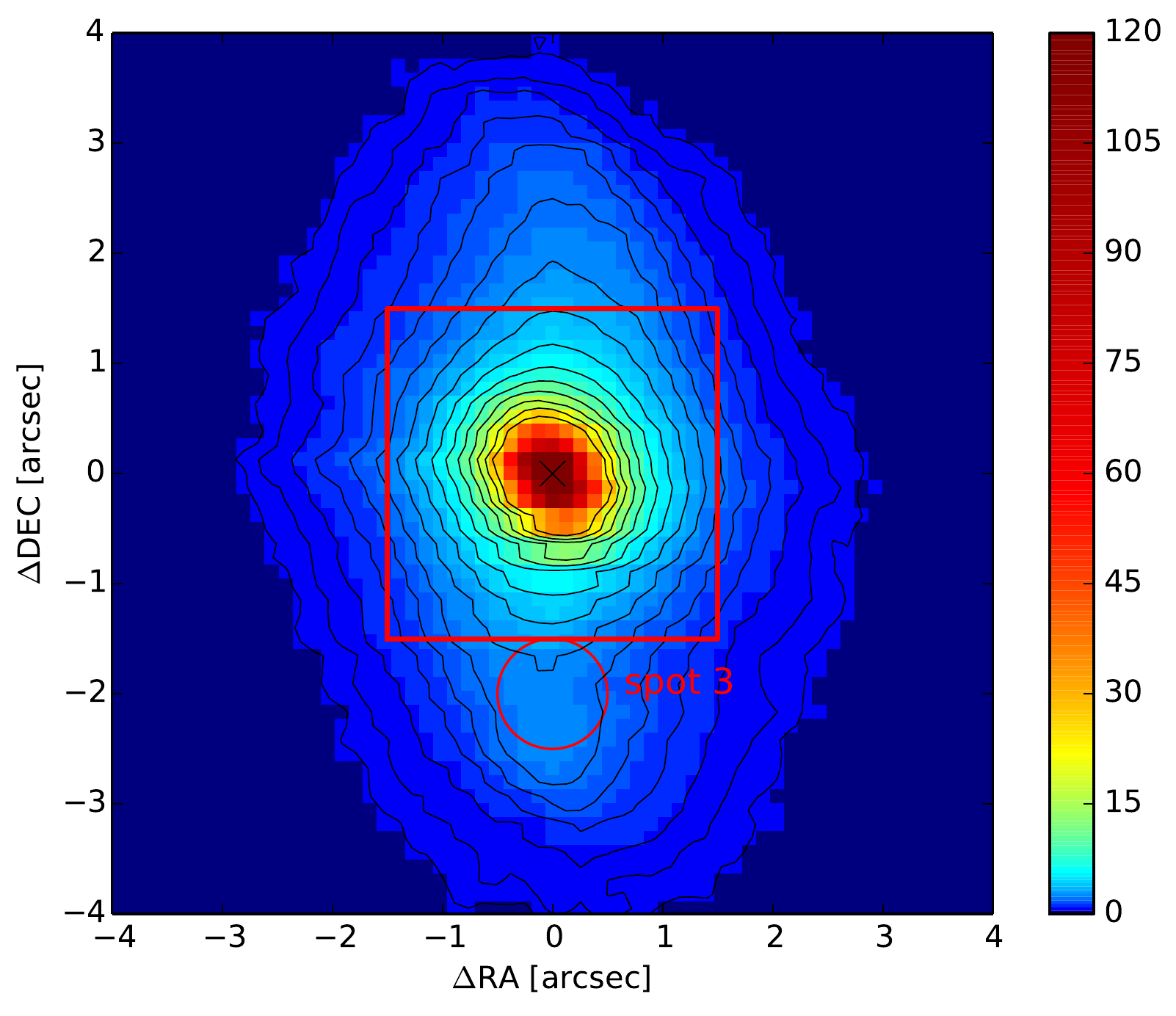}
\includegraphics[width=0.45\linewidth]{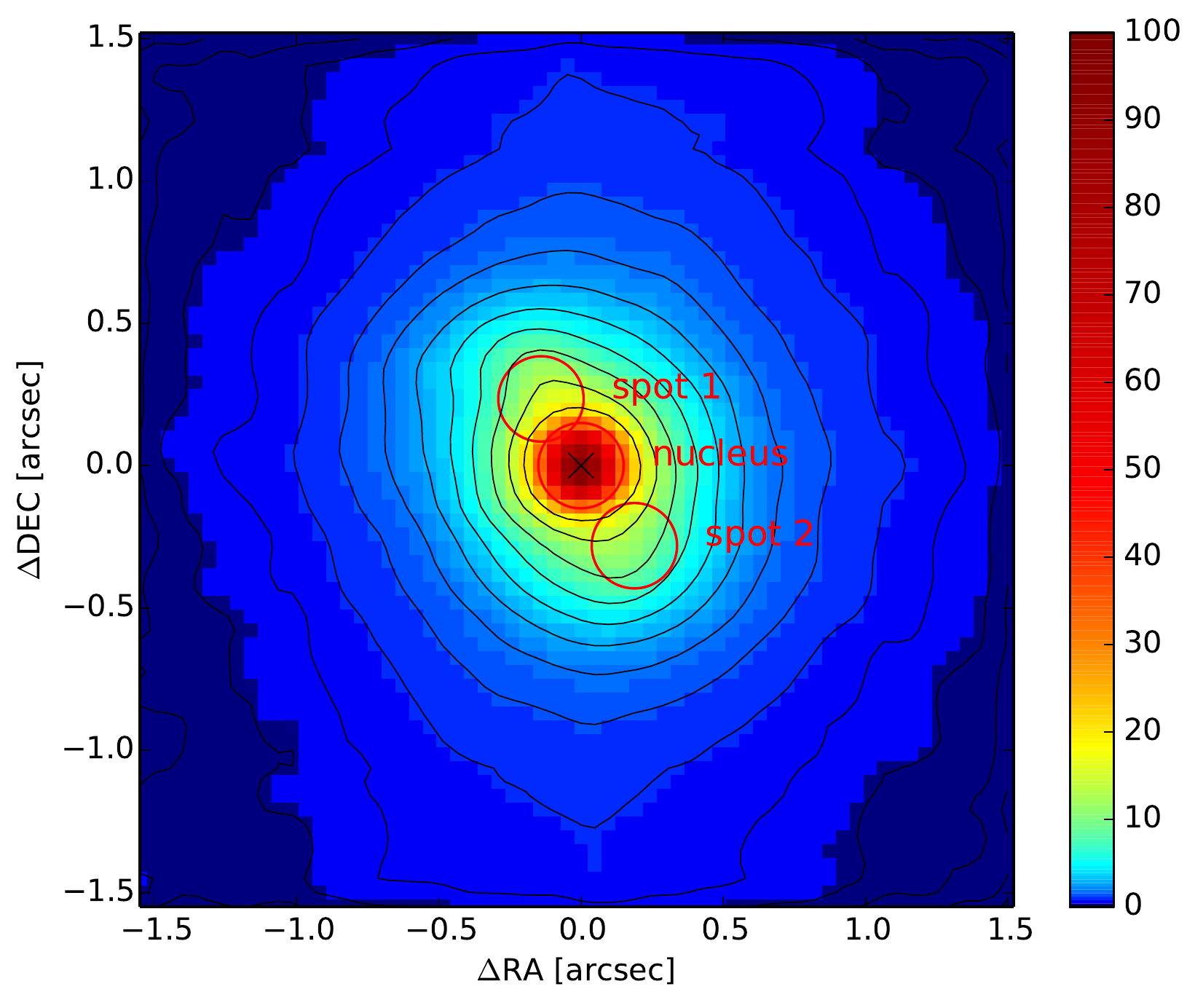}\\
\includegraphics[width=\linewidth]{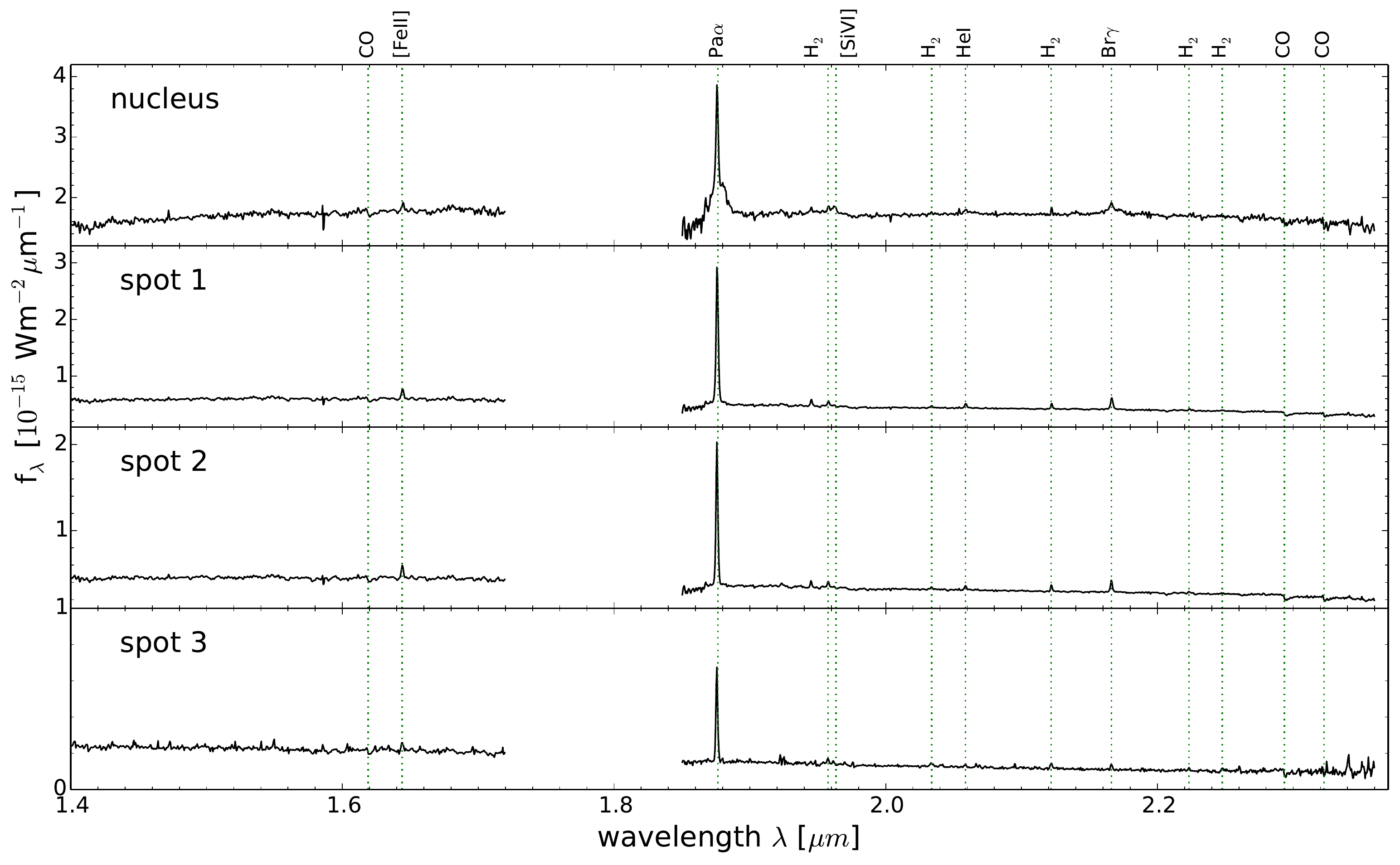}
\caption{In the \emph{upper panels} we show the continuum emission around $2.2\,\mu\mathrm{m}$ in the $8\arcsec \times 8\arcsec$ and the $3\arcsec \times 3\arcsec$ FOV, in unit $10^{-18}\,\mathrm{W}\,\mathrm{m}^{-2}\,\mu\mathrm{m}^{-1}$. In the \emph{bottom panels}, spectra extracted from three apertures with radius $0\farcs15$ in the $3\arcsec \times 3\arcsec$-FOV and from one aperture with radius $0\farcs5$ in the $8\arcsec \times 8\arcsec$-FOV that are marked in the continuum images are shown. The emission lines and the CO absorption band heads are marked in the spectra.}
\label{fig:spotspectra}
\end{figure*}

The spectra obtained with SINFONI show a big variety of emission lines. In the top right panel of Fig.~\ref{fig:spotspectra}, we mark three spots, the nuclear region and two off-nuclear regions with a radius of $0\farcs15$ each. Furthermore, we extract one spectrum from an aperture $2\arcsec$ south of the nucleus. Here we chose a larger aperture with a radius of $0.5\arcsec$ matching the lower spatial resolution in the $8\arcsec \times 8\arcsec$ data set. The maps show the continuum emission around $2.2\,\mu\mathrm{m}$ from the SINFONI-cube, left for the $8\arcsec \times 8 \arcsec$ and right for the $3\arcsec \times 3\arcsec$ FOV. In the lower panels, we show three spectra that are extracted from these regions. In the nuclear spectrum, broad line components and the coronal line [\ion{Si}{vi}] are apparent, which indicate the presence of strong radiation from the central AGN. In the off-nuclear spectra, stellar features like the CO band heads become more prominent. Moreover, 3D-spectroscopy enables us to extract spectra and fit the emission lines in every spectral pixel (spaxel), which results in maps of the emission line fluxes.

\begin{figure*}
\centering
\includegraphics[width=0.3\linewidth]{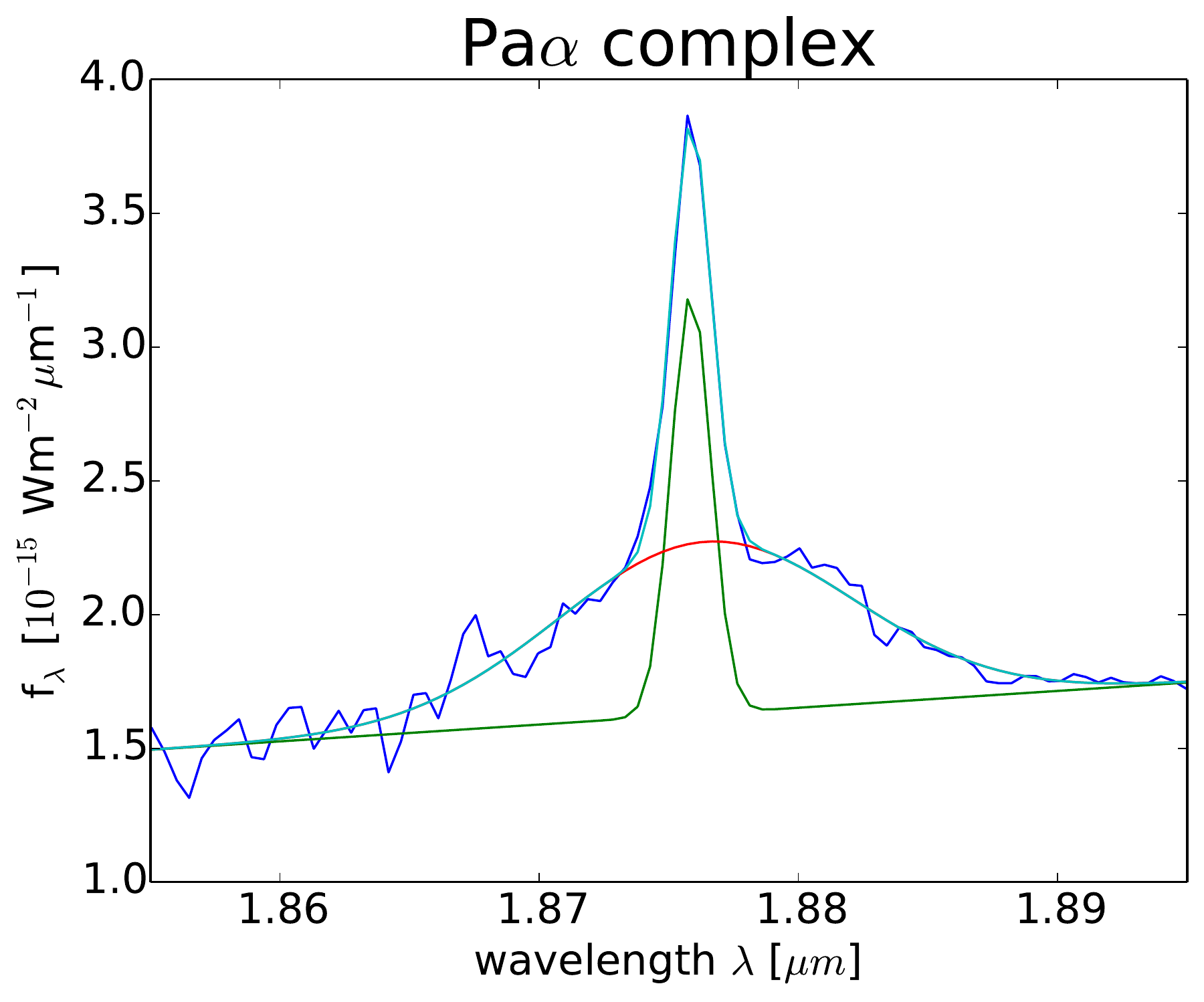}
\includegraphics[width=0.3\linewidth]{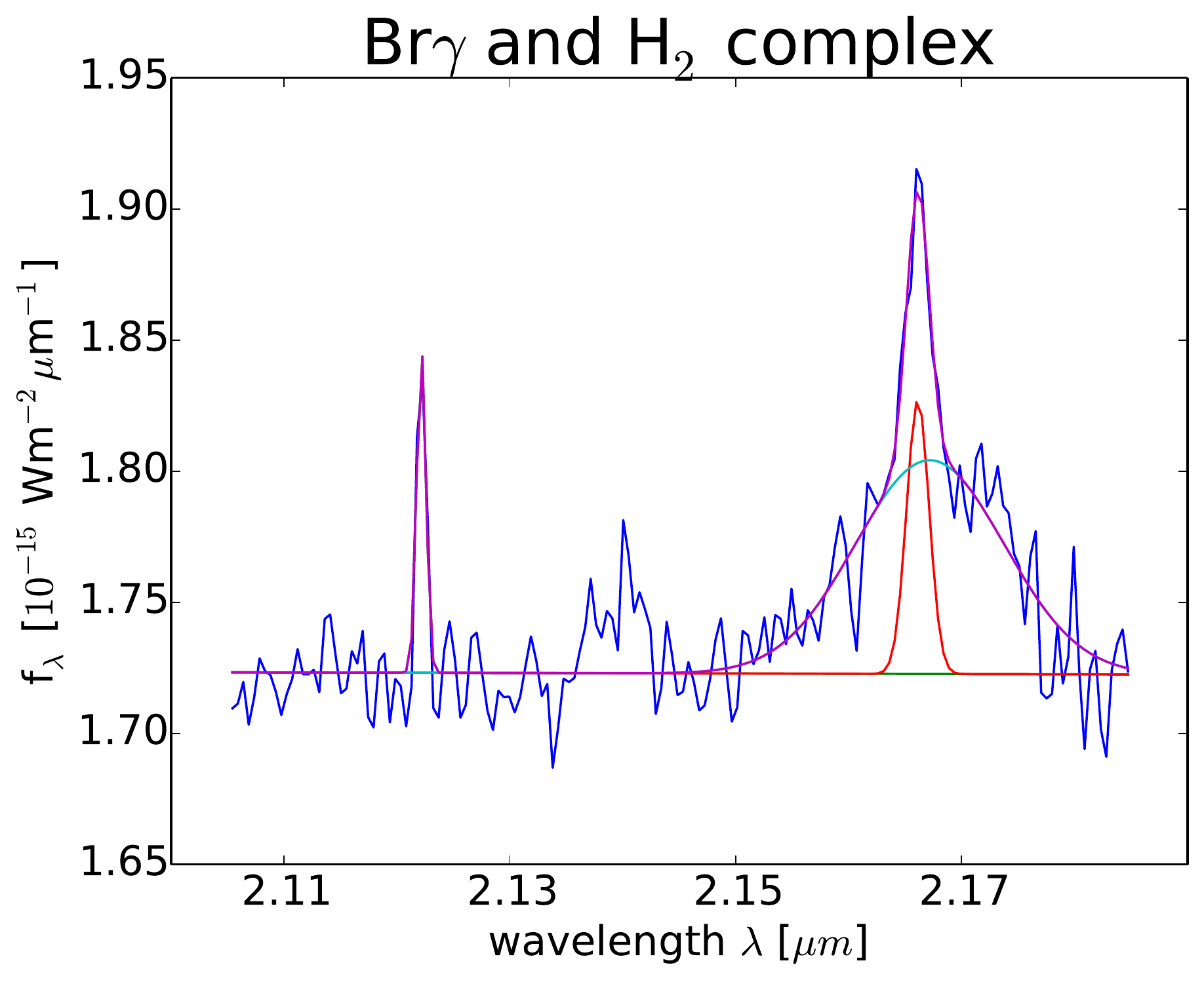}
\includegraphics[width=0.3\linewidth]{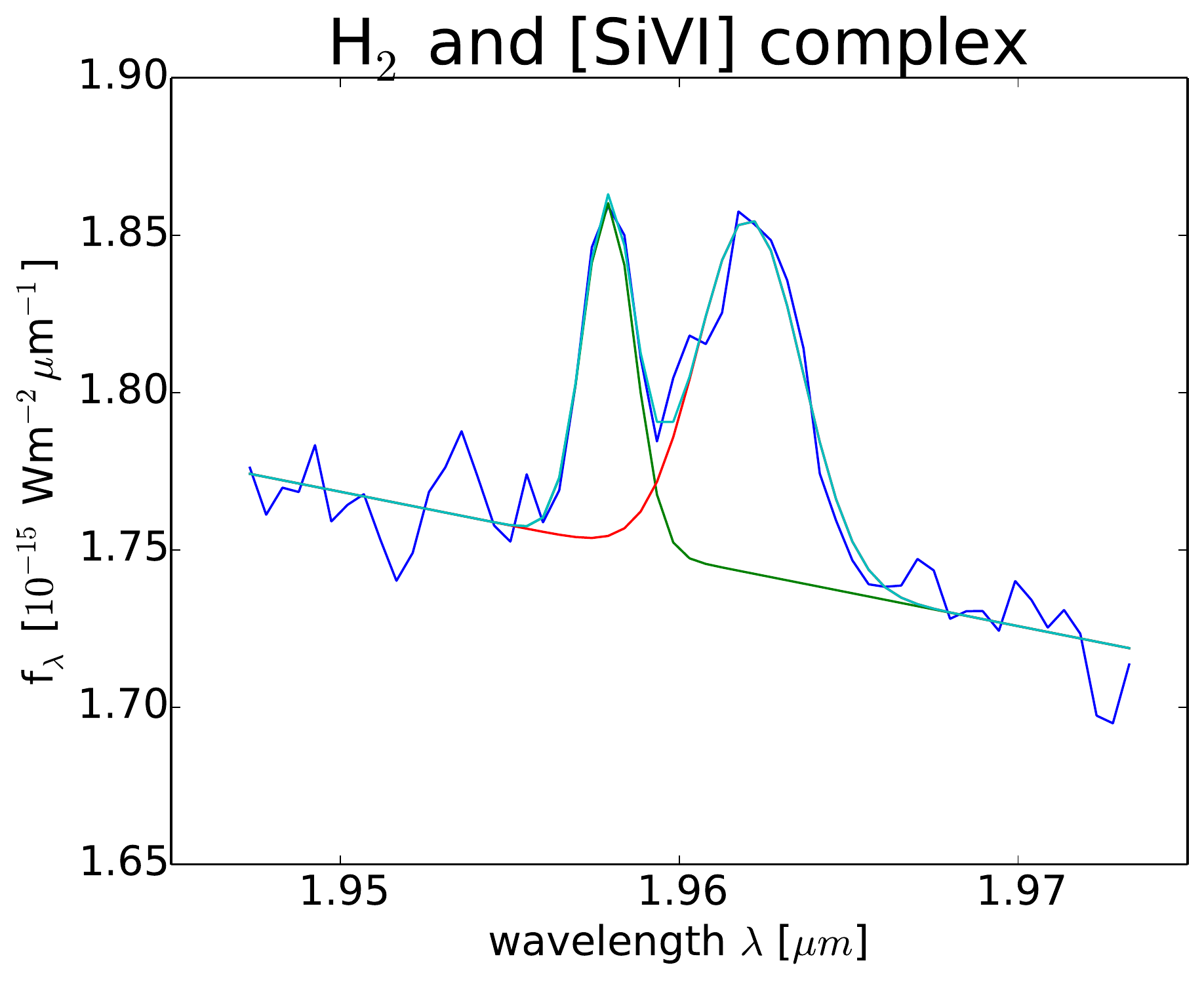}
\caption{Three examples for the fits applied to the cubes. The plots show the fits of the corresponding complex in the nuclear aperture.}
\label{fig:gaussfits}
\end{figure*}

All emission lines have been fitted with a single Gaussian component plus zero-order polynomial. In case of the hydrogen recombination lines Pa$\alpha$ and Br$\gamma$, a second component, accounting for the light emitted from the broad line region (BLR) was added. Emission lines that are close to each other (e.g., [\ion{Si}{vi}] and H$_2 \lambda$1.96$\mu$m) have been fitted simultaneously. Examples for these fits are displayed in Fig.~\ref{fig:gaussfits}. We use the \textsc{python}-implementation of \textsc{mpfitexpr} \citep{2009ASPC..411..251M}. Errors of the fit parameters are calculated from the covariance matrix. For testing purposes, we also determined uncertainties by generating 100 Monte Carlo realizations of the input spectrum, adding Gaussian noise with a width corresponding to the standard deviation in a line-free region. Both methods deliver comparable results. If not noted otherwise, we clipped values in the maps with a relative error larger than 30\%.

In Table~\ref{tab:emline-fluxes}, we present the flux measurements for the emission lines in three apertures of radius $0\farcs15$ each that are marked in the continuum image in Fig.~\ref{fig:spotspectra}.

Two strong hydrogen recombination lines were detected: Br$\gamma$ and Pa$\alpha$. In the central region, both show two spectral components, a broad and a narrow one. The broad components show a FWHM of $\approx 2000\,\mathrm{km}\,\mathrm{s}^{-1}$, indicative for a type-1 AGN. According to the unified model \citep[e.g.,][]{1993ARA&A..31..473A,1995PASP..107..803U}, this emission stems from clouds very close \citep[$r\sim $lightdays, see review of][]{1993PASP..105..247P} to the central supermassive black hole which are moving at high velocities.

\begin{figure*}
\includegraphics[width=0.45\linewidth]{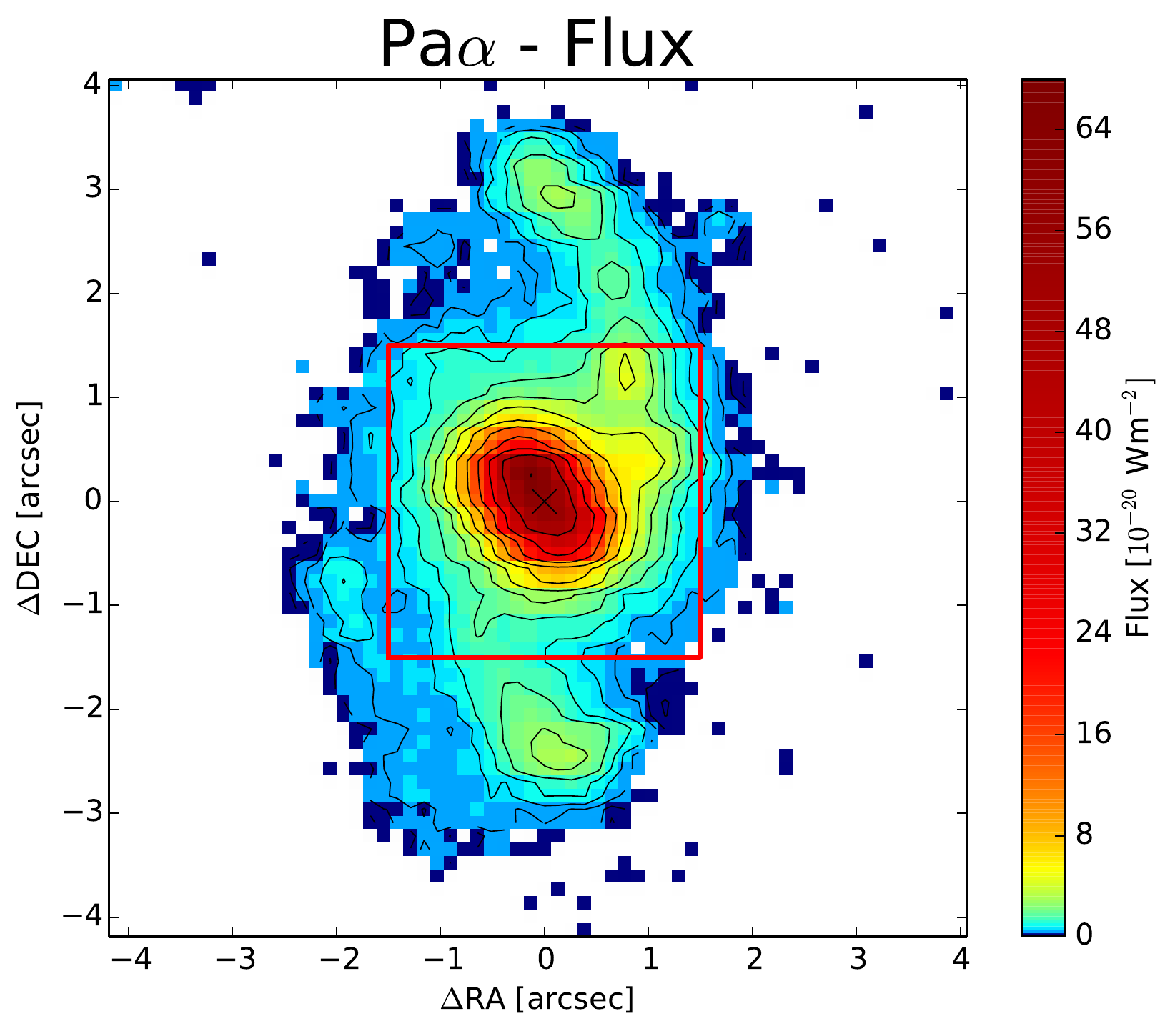}
\includegraphics[width=0.45\linewidth]{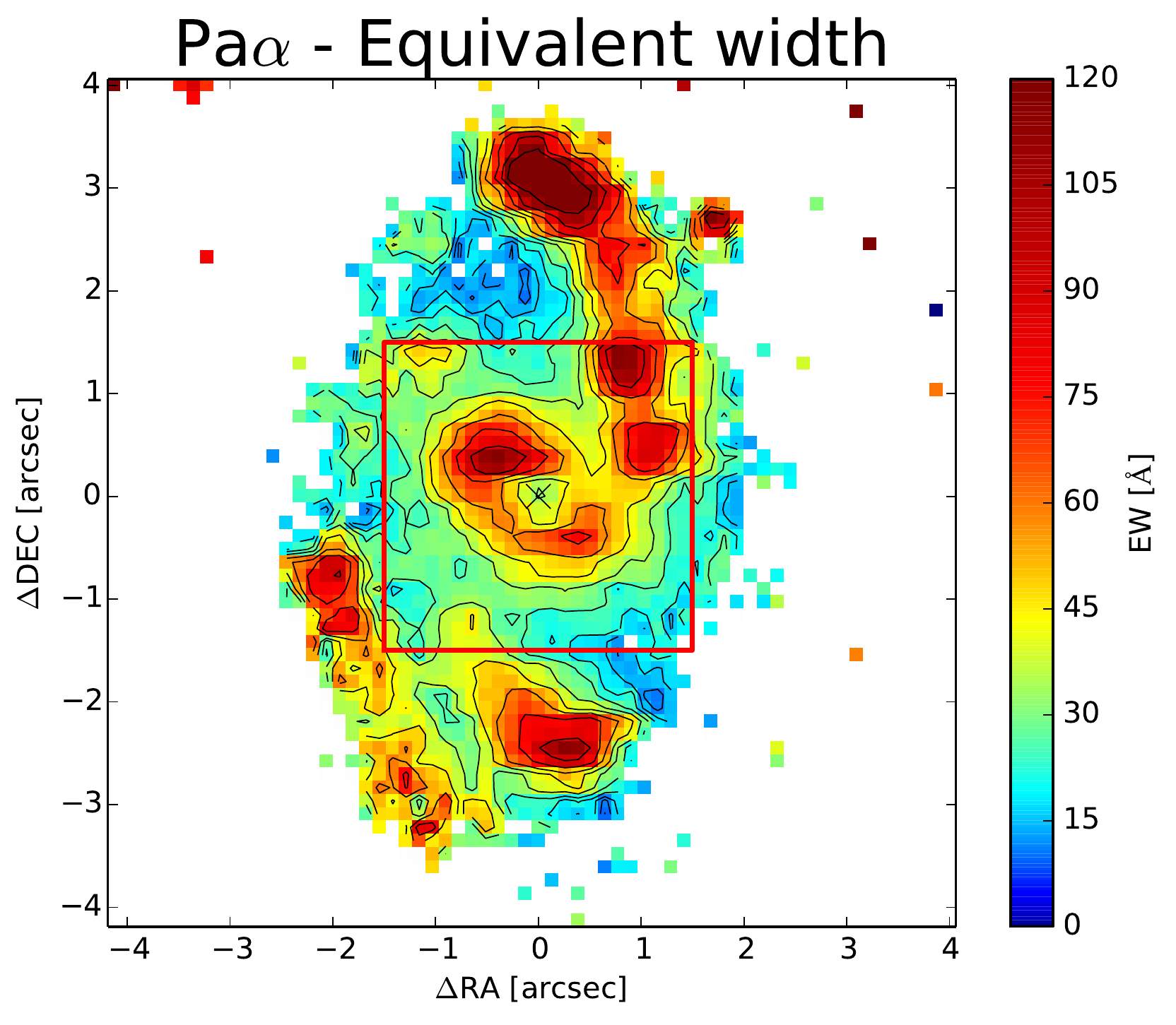}

\includegraphics[width=0.45\linewidth]{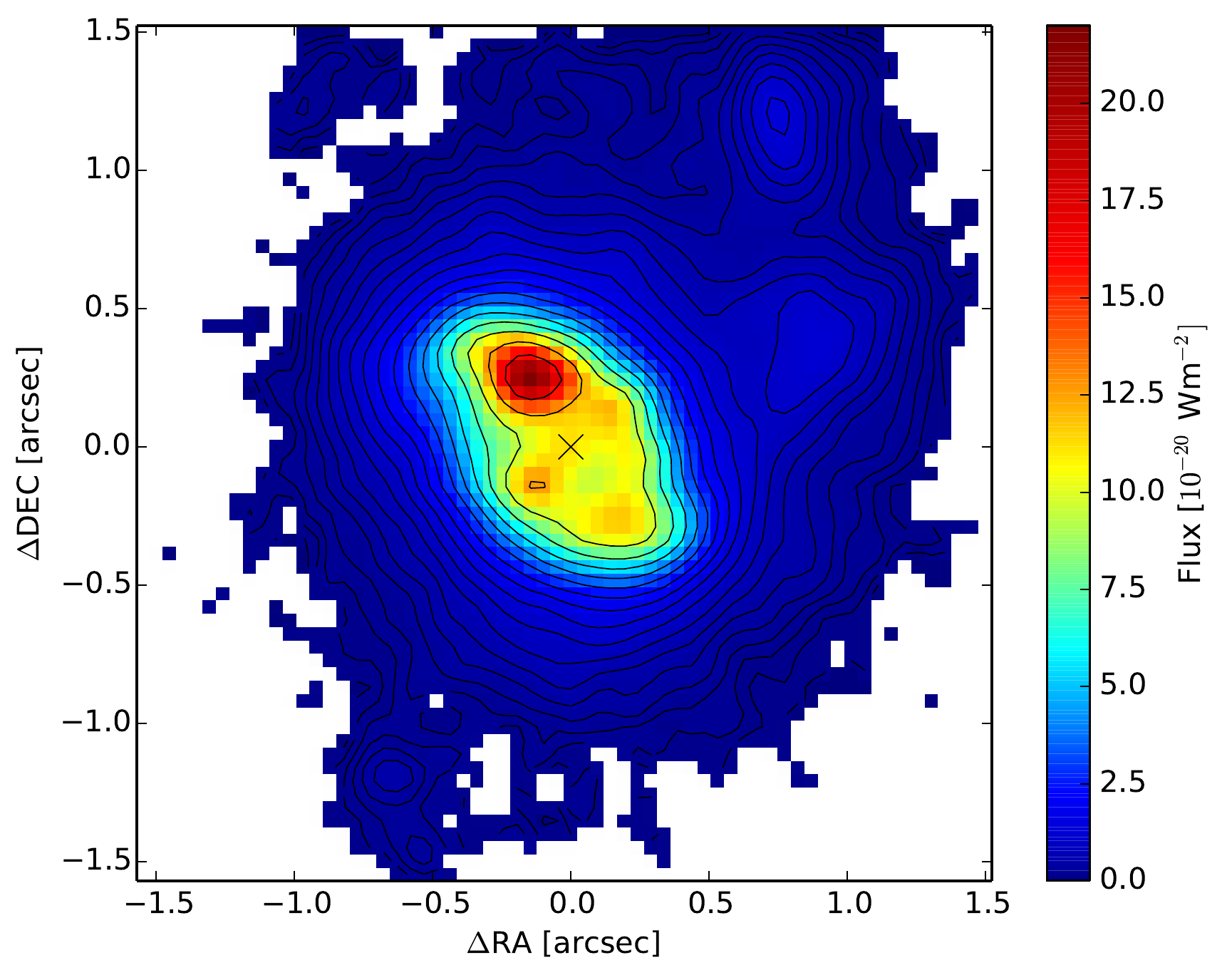}
\includegraphics[width=0.45\linewidth]{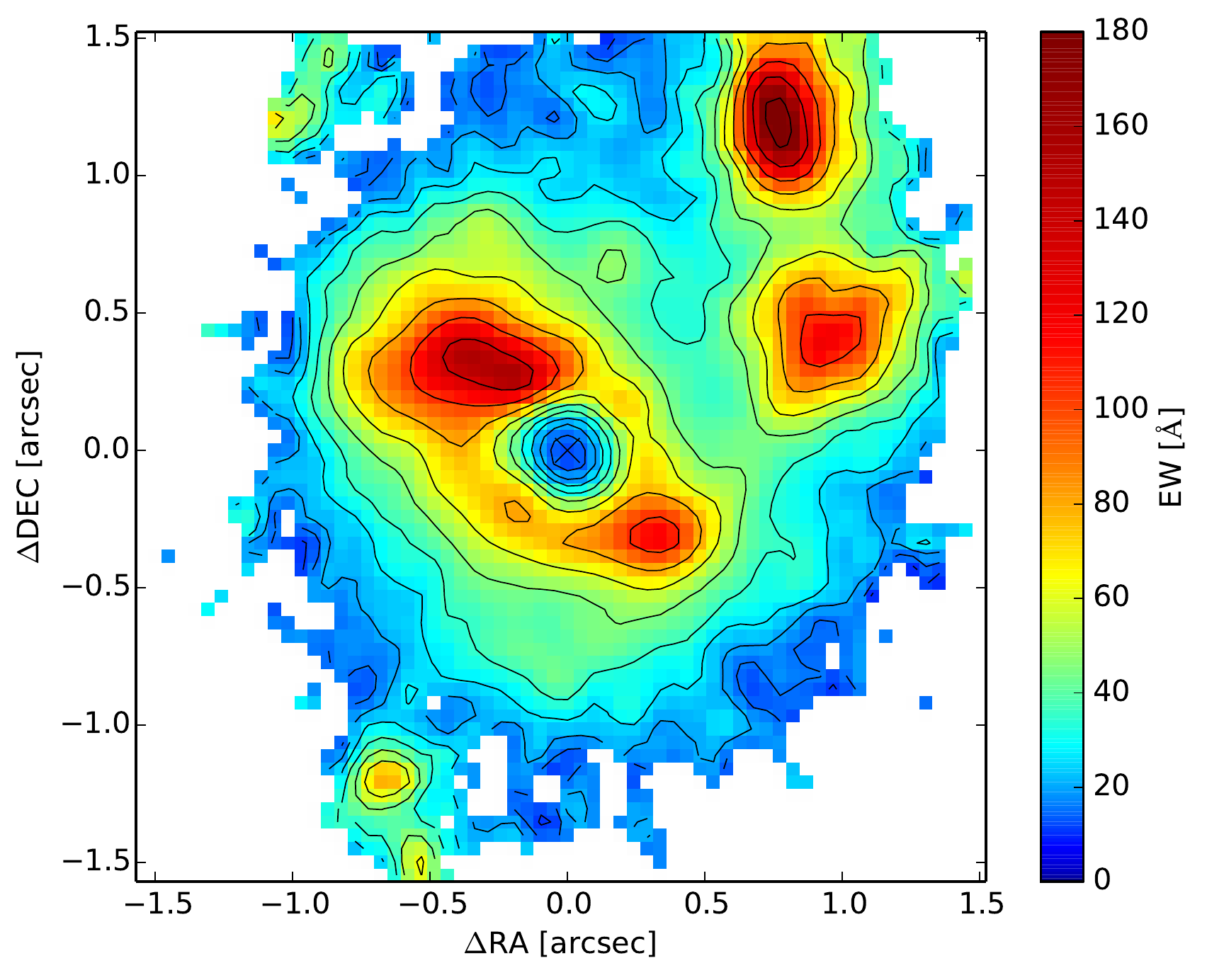}
\caption{\emph{From left to right:} Flux and equivalent width of the narrow component of the hydrogen recombination line Pa$\alpha$. \emph{Upper row:} FOV $8\arcsec \times 8\arcsec$, \emph{lower row:} FOV $3\arcsec \times 3\arcsec$. 
The center of the continuum emission is marked by a $\times$.}
\label{fig:paa}
\end{figure*}

\begin{figure*}
\includegraphics[width=0.3\linewidth]{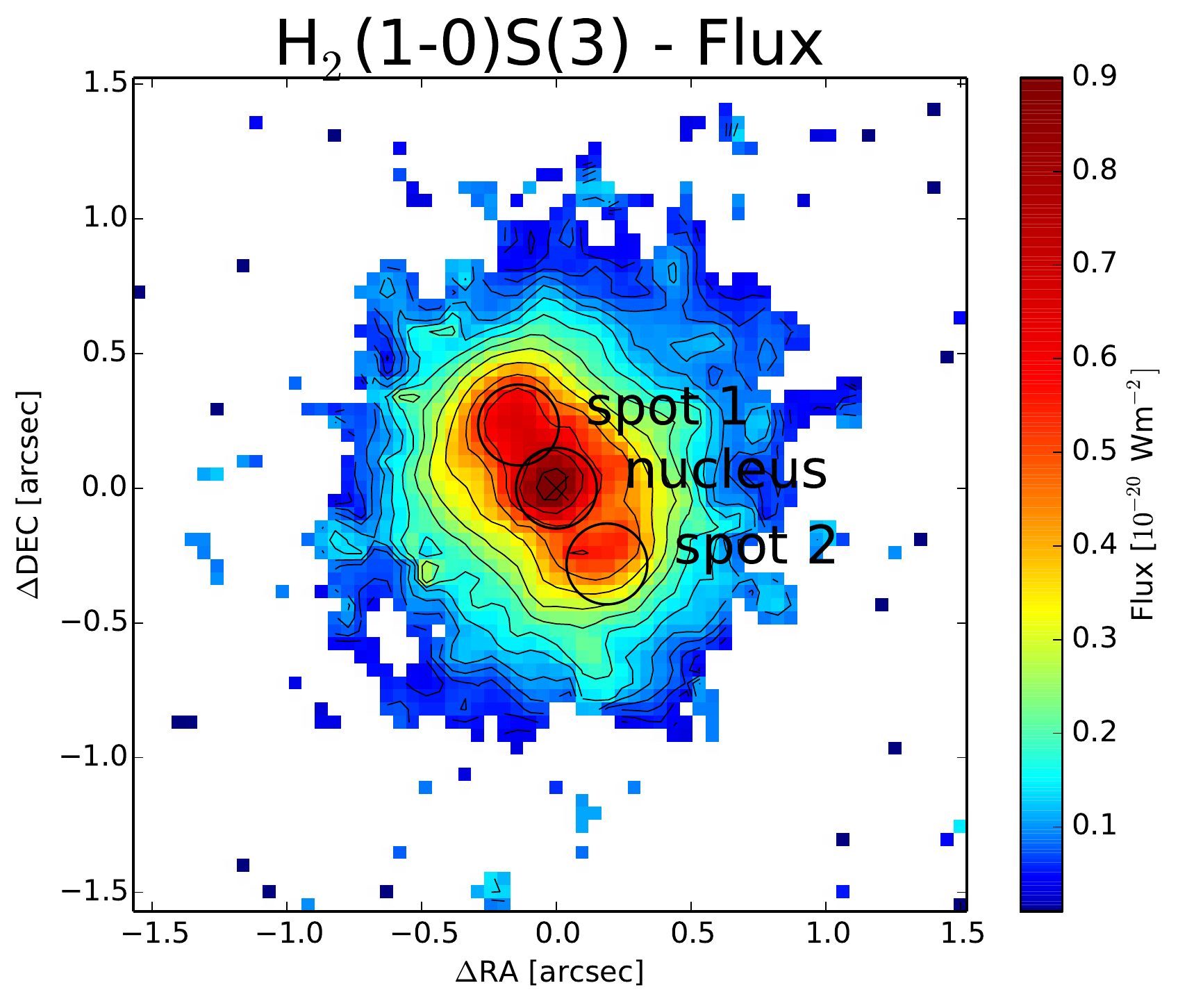}
\includegraphics[width=0.3\linewidth]{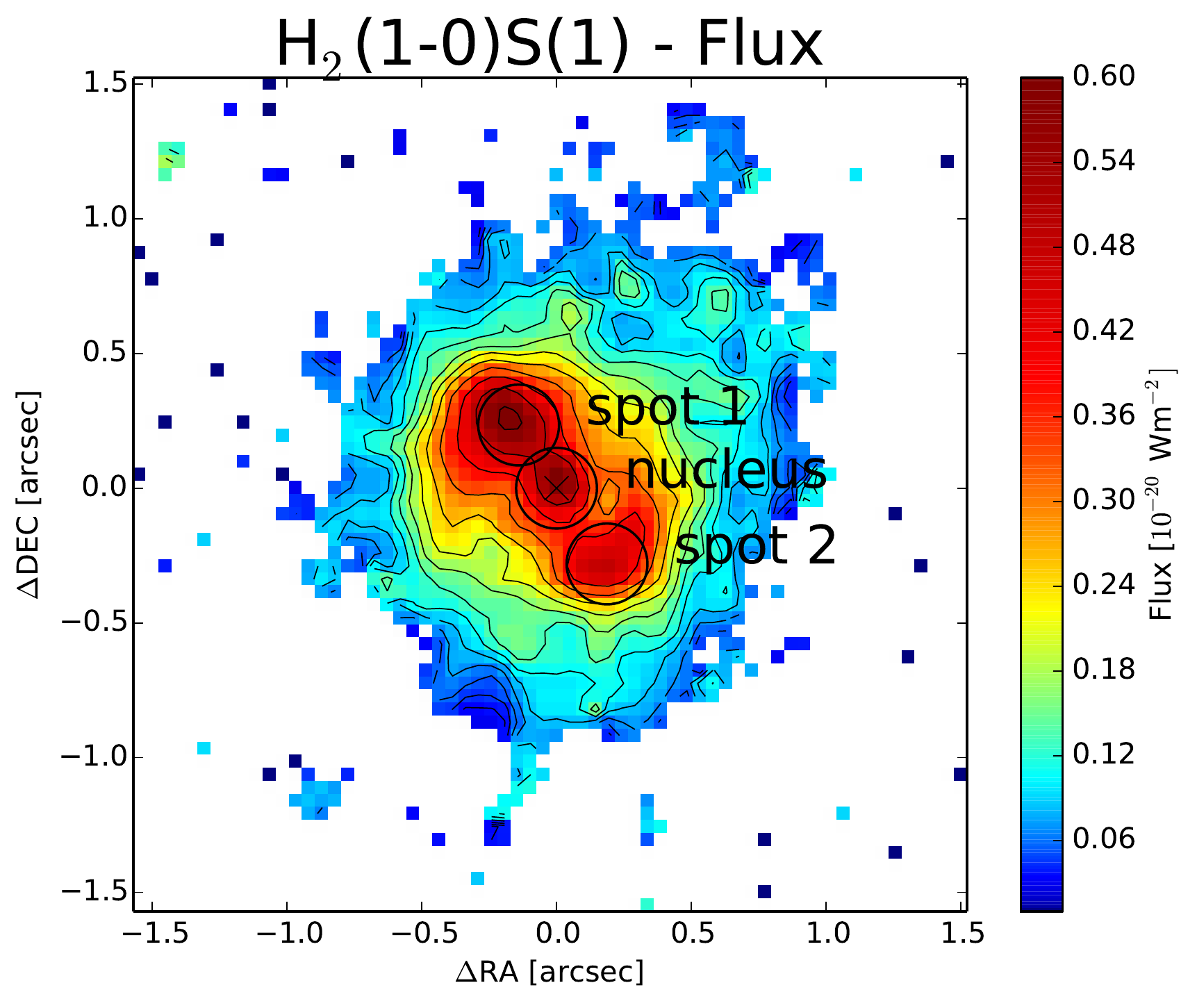}
\includegraphics[width=0.3\linewidth]{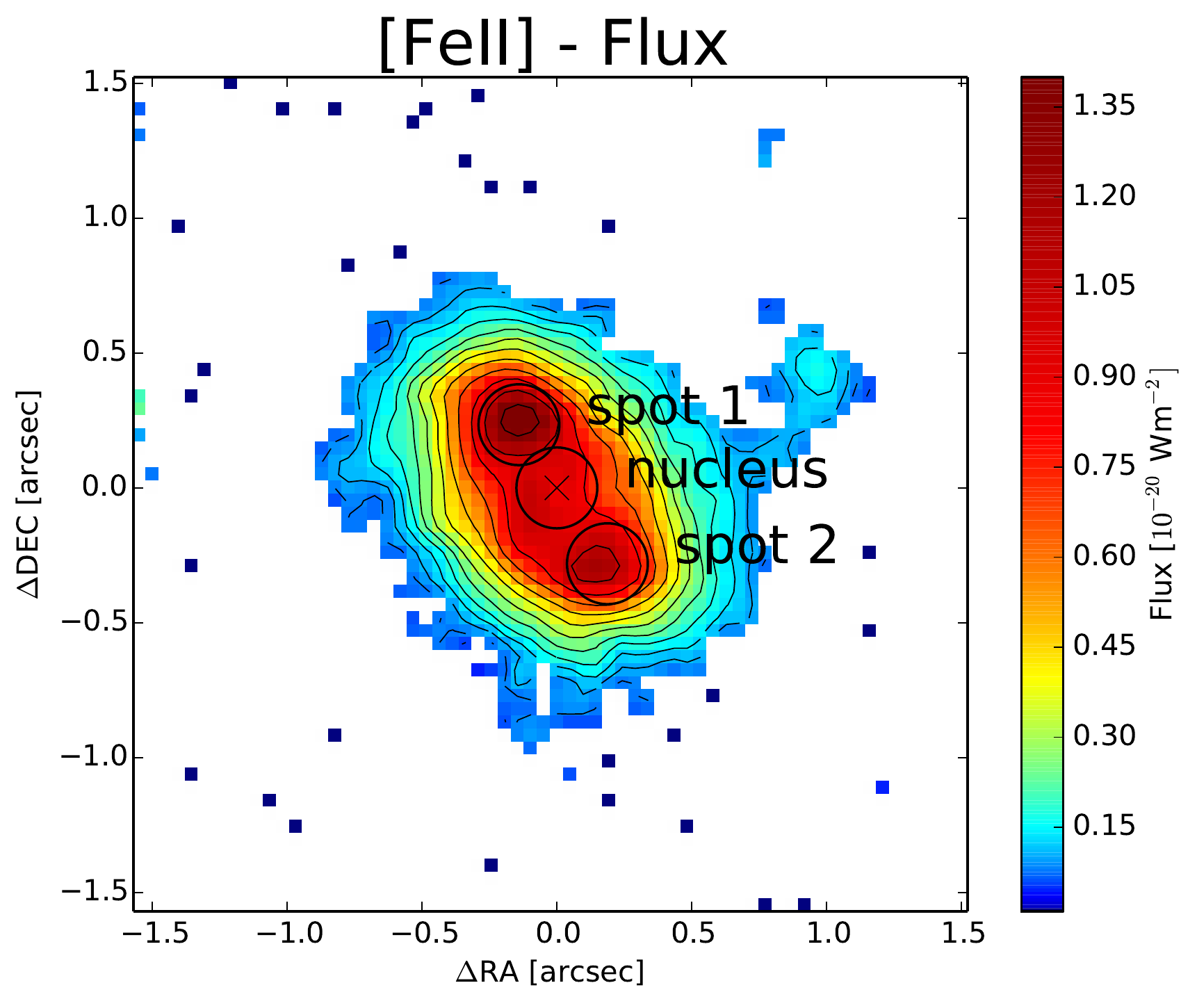}
\caption{Flux of (\emph{from left to right}) H$_2$(1-0)S(3) $\lambda$1.96 $\mu$m, H$_2$(1-0)S(1) $\lambda$2.12 $\mu$m, and [\ion{Fe}{ii}] $\lambda$1.64 $\mu$m. The center of the continuum emission is marked by a $\times$. The three apertures nucleus, spot1, and spot2 are marked as well.}
\label{fig:h2maps}
\end{figure*}

HE 1029-1831 shows extended narrow emission in Br$\gamma$ and Pa$\alpha$. In the Pa$\alpha$ emission two spiral arms in the ionized gas emission are clearly visible (Fig. \ref{fig:paa}). Since Br$\gamma$ is less strong, the arms are not as prominent (Fig. \ref{fig:brg}). 
The AO-assisted $3\arcsec \times 3\arcsec$-observations reveal a patchy ring in ionized gas emission of about $0\farcs3=240\,\mathrm{pc}$ in radius and an ellipticity of $\epsilon=0.29$. This is in good agreement with the range of values that \cite{2008AJ....135..479B} find in a SINFONI study of the circumnuclear starforming rings of five nearby spiral galaxies. At the points where the gas spirals meet the ring, the gas emission is particularly high. The [\ion{Fe}{ii}]-emission does not peak in the center but in two off-nuclear regions that are identical with the two off-nuclear Pa$\alpha$/Br$\gamma$ peaks (Fig. \ref{fig:h2maps}). Thus, we conclude that the strong circumnuclear gas emission does not mainly stem from the narrow-line region but traces a circumnuclear star-formation region. 

Furthermore, we see emission from the coronal line [\ion{Si}{vi}]. We fit a Gaussian function to the spatial distribution and get a FWHM of $0\farcs2$ for the $3\arcsec \times 3\arcsec$-FOV and $0\farcs5$ for the $8\arcsec \times 8\arcsec$ FOV. Coronal lines are forbidden transitions in the ground levels of highly ionized atoms that have ionization potentials above $100\,\mathrm{eV}$. Thus, they are free of contribution from star formation and directly associated with the AGN. \cite{2006A&A...454..481M} argue that coronal line emission stems from the inner NLR and that they are probably associated with outflows. Recently, \cite{2011ApJ...739...69M} and \cite{2013MNRAS.430.2411M} observed extended emission in the coronal lines of NGC 1068 as well as a great complexity in the morphology and kinematics. This indicates that the nature of coronal line emission is far from being understood.

For the broad components of Pa$\alpha$ and Br$\gamma$, we find a FWHM of $0\farcs2$ for the $3\arcsec \times 3\arcsec$-FOV and $0\farcs5 - 0\farcs6$ for the $8\arcsec \times 8\arcsec$-FOV. 
For the $8\arcsec \times 8\arcsec$ FOV, these values agree with the values derived from the telluric star observations in Sect. \ref{sec:obs_sinfo}. For the $3\arcsec \times 3\arcsec$-FOV, the values derived here are slightly higher. We claim that they are a more robust method to derive the spatial resolution since they are extracted from the analyzed data itself.

Several $\mathrm{H}_2$ molecular hydrogen emission lines are detected: H$_2$(1-0)S(3) $\lambda$1.96 $\mu$m, H$_2$(1-0)S(2) $\lambda$2.03 $\mu$m, H$_2$(1-0)S(1) $\lambda$2.12 $\mu$m, H$_2$(1-0)S(0) $\lambda$2.22 $\mu$m, and H$_2$(2-1)S(1) $\lambda$2.25 $\mu$m. Other lines that are visible are \ion{He}{i} as well as several CO stellar absorption bands.

\begin{table*}
\centering
\caption{Emission-line fluxes for three positions, the center and two off-center spots. Apertures have a radius of $0\farcs 15$. } 
\label{tab:emline-fluxes}
\begin{tabular}{cccccccc} \hline \hline
 & & \multicolumn{2}{c}{nucleus} & \multicolumn{2}{c}{spot 1} & \multicolumn{2}{c}{spot 2} \\ 
line & $\lambda$[$\mu$m] & Flux & FWHM & Flux & FWHM & Flux & FWHM \\ \hline

[\ion{Fe}{ii}] & 1.64400 & $26 \pm 3$ & $350$ & $34.1 \pm 1.9$ & $340$ & $27.1 \pm 1.8$ & $310$ \\
Pa$\alpha$ (narrow) & 1.87561 & $304 \pm 13$ & $290$ & $442 \pm 5$ & $280$ & $276 \pm 3$ & $260$ \\
Pa$\alpha$ (broad) & 1.87561 & $920 \pm 39$ & $2150$\tablefootmark{*} & $186 \pm 15$ & $2150$ & $92 \pm 10$ & $2150$\tablefootmark{*} \\
H$_2$(1-0)S(3) & 1.95756 & $20.9 \pm 1.9$ & $280$ & $16.8 \pm 1.7$ & $270$ & $12.8 \pm 1.2$ & $260$ \\

[\ion{Si}{vi}] & 1.9634 & $43 \pm 3$ & $560$ & $0.0 \pm 0.0$ & $290$ & $3.3 \pm 1.5$ & $350$ \\
H$_2$(1-0)S(2) & 2.03376 & $5.6 \pm 1.6$ & $180$ & $6.0 \pm 0.7$ & $250$ & $4.2 \pm 0.6$ & $210$ \\

\ion{He}{i} & 2.05869 & $10 \pm 2$ & $270$ & $14.9 \pm 1.0$ & $260$ & $7.7 \pm 0.8$ & $230$ \\
H$_2$(1-0)S(1) & 2.12183 & $12.7 \pm 1.7$ & $140$ & $14.4 \pm 0.8$ & $200$ & $11.1 \pm 0.5$ & $190$ \\
Br$\gamma$ (narrow) & 2.16612 & $26 \pm 4$ & $330$ & $37.8 \pm 1.1$ & $260$ & $22.0 \pm 0.7$ & $220$ \\
Br$\gamma$ (broad) & 2.16612 & $135 \pm 11$ & $2150$\tablefootmark{*} & $25 \pm 4$ & $2150$ & $14 \pm 3$ & $2150$\tablefootmark{*} \\
H$_2$(1-0)S(0) & 2.22329 & $5 \pm 3$ & $150$ & $5.4 \pm 1.0$ & $230$ & $4.5 \pm 1.0$ & $280$ \\
H$_2$(2-1)S(1) & 2.24772 & --- & --- & $4.3 \pm 0.9$ & $370$ & $2.7 \pm 0.7$ & $350$ \\

\hline
\end{tabular}
\tablefoot{All flux measurements are in units of $10^{-20}\,\mathrm{W}\,\mathrm{m}^{-2}$. FWHMs are in $\mathrm{km}\,\mathrm{s}^{-1}$ and not corrected for instrumental broadening. \tablefoottext{*}{Broad line width was fixed to the line width measured for the broad Pa$\alpha$ component in the nuclear aperture.}}
\end{table*}

\subsection{Stellar continuum}
\label{sec:stellar}

\begin{figure*}
\sidecaption
\includegraphics[width=12cm]{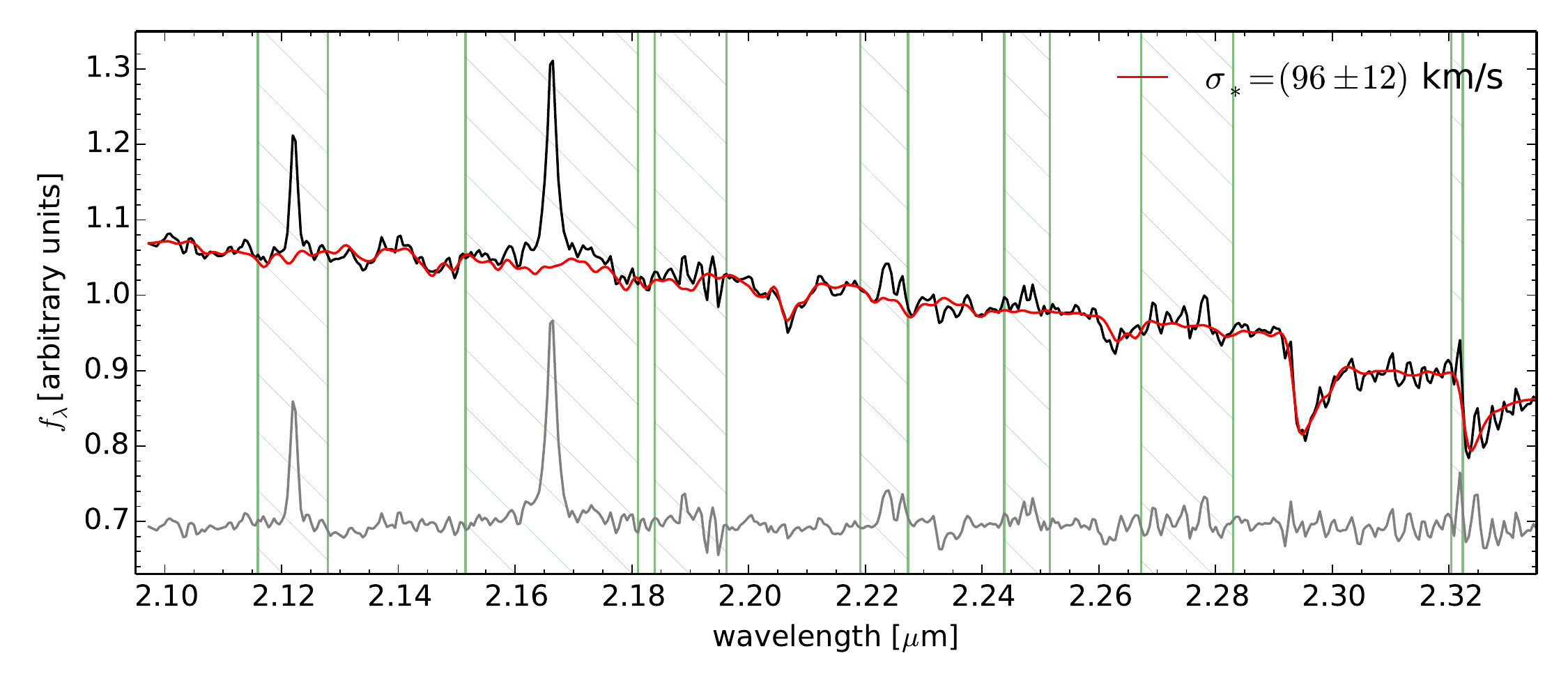}
\caption{Fit of the stellar kinematics of the $K$-band spectrum of HE 1029-1831. The radius of the aperture is $r=0.6\arcsec=470\,\mathrm{pc}$, corresponding to the effective radius. The spectrum is shown in black, the best fitting combination of NIFS templates in red. Spectral regions that are contaminated by emission lines or telluric features are masked and are marked by the shaded areas. The residuum, shifted vertically for clarity, is shown in gray.}
\label{fig:ppxffit}
\end{figure*}

Under the assumption that the central bulge component is supported by random motion rather than ordered rotation, the stellar velocity dispersion can be used for virial mass estimates and black hole estimates via the $M_\mathrm{BH}-\sigma_*$ relation. To measure the stellar velocity dispersion $\sigma_*$, we fit stellar templates to the $K$-band spectrum that we integrated over an aperture corresponding to the effective radius measured in Sect.~\ref{sec:decomp}. We use the penalized pixel-fitting algorithm \citep[pPXF,][]{2004PASP..116..138C} and take stellar templates from the Gemini Spectral Library of Near-IR Late-Type Stellar Templates \citep{2009ApJS..185..186W} which consists of 11 giant and supergiant stars with spectral classes from G8 to M5 that have been observed with the GEMINI integral-field spectrograph NIFS. The $H+K$ grating of SINFONI has a spectral resolution of $R=1500$ at a wavelength of around $\lambda=1.9\,\mu\mathrm{m}$ which we confirmed by measuring the width of OH-lines. The stellar templates have a different spectral resolution. Therefore, the templates had to be degraded down to the resolution of SINFONI first. Regions that contain emission lines or telluric features are masked. Fig.~\ref{fig:ppxffit} shows the best fit with a stellar velocity dispersion of $\sigma_*=(96\pm 12)\,\mathrm{km}\,\mathrm{s}^{-1}$. To test our results, we also use a set of theoretical spectra of red giant and supergiant stars from \cite{2007A&A...468..205L} and get a consistent result $\sigma_*=(111\pm 14)\,\mathrm{km}\,\mathrm{s}^{-1}$. In the following sections, we will use a mean value of $\sigma_*=104\,\mathrm{km}\,\mathrm{s}^{-1}$. To estimate the uncertainties, we perform a Monte Carlo simulation. We generate 1000 spectra by adding Gaussian noise to the original spectrum. The width of the normal distribution of the noise is given by the standard deviation of the flux in a line-free region. The uncertainty is then given by the standard deviation of the 1000 fit results.

\subsection{Extinction}
\label{sec:ext}

Extinction can have a significant effect on the line fluxes. Therefore, we use hydrogen recombination line ratios to estimate the extinction. We follow the extinction law derived by \cite{1989ApJ...345..245C} to estimate the extinction in the near-infrared:

\begin{equation}
A_V = 94.6 \times \log \left( \frac{ ( f_{\mathrm{Br}\gamma}/f_{\mathrm{Pa}\alpha} )_\mathrm{obs}}{( f_{\mathrm{Br}\gamma}/f_{\mathrm{Pa}\alpha} )_\mathrm{intr}} \right)
\end{equation}
where $( f_{\mathrm{Br}\gamma}/f_{\mathrm{Pa}\alpha} )_\mathrm{obs}$ is the observed line ratio between Br$\gamma$ and Pa$\alpha$ that is divided by the intrinsic line ratio $( f_{\mathrm{Br}\gamma}/f_{\mathrm{Pa}\alpha} )_\mathrm{intr}=0.083$ taken from \cite{2006agna.book.....O} assuming case B recombination with a typical electron density of $n_e = 10^4\,\mathrm{cm}^{-3}$ and a temperature $T = 10^4\,\mathrm{K}$. In the nuclear aperture, we measure an extinction of $A_V=1.6$ from the narrow lines 
For the two other apertures, we get values of $A_V=1.2$ (spot 1) and $A_V=-1.6$ (spot 2). 
The uncertainties of $A_V$ are up to 100\%. Thus, we decide not to correct the near-infrared fluxes for extinction. However, we have to keep in mind that this introduces an error. In particular, ignoring an extinction of $A_V=2$ will cause a loss in flux of around $30\%$.

\section{Discussion}
\label{sec:discussion}

\subsection{The central black hole}
\label{sec:bh}

The observed correlations between black hole mass and host galaxy properties imply that the evolution of the central supermassive black holes and their hosts might be connected to each other. Black hole masses are thus of great interest. From single epoch spectroscopy, we expect a black hole mass of around $\log(M_\mathrm{BH}/M_\odot)=7.3-7.4$. We compare this mass to other mass estimates (Table \ref{tab:bhmasses}) and find an Eddington ratio of $\lambda \approx 0.3$, implying that the black hole is efficiently growing.

\begin{table}
\centering
\caption{Black hole mass estimates from different relations.} 
\label{tab:bhmasses}
\begin{tabular}{cc} \hline \hline
Used relation & $\log(M_\mathrm{BH}/M_\odot)$ \\ \hline
\multicolumn{2}{c}{\emph{broad line width}} \\
A.~Schulze (priv.~comm.) & 7.4 \\
J.~Scharw\"achter (priv.~comm.) & 7.3 \\
\cite{2010ApJ...724..386K} & 7.3 \\

\multicolumn{2}{c}{\emph{$M_\mathrm{BH}-\sigma_*$ relation}} \\
\cite{2009ApJ...698..198G} & 7.0 (+0.5\tablefootmark{a})\\
\cite{2011MNRAS.412.2211G} & 6.7\tablefootmark{b} (+0.5\tablefootmark{a}) \\
Kormendy \& Ho (2013) & 7.4 (+0.5\tablefootmark{a})\\

\multicolumn{2}{c}{\emph{$M_\mathrm{BH}-L_\mathrm{bulge}$ relation}\tablefootmark{c}} \\
\cite{2003ApJ...589L..21M} & 8.0 \\
\cite{2012MNRAS.419.2264V} & 8.2 \\
\cite{2013ApJ...764..151G} & 8.4 \\
Kormendy \& Ho (2013) & 8.4 \\

\multicolumn{2}{c}{\emph{$M_\mathrm{BH}-M_\mathrm{bulge}$ relation}} \\
\cite{2004ApJ...604L..89H} & 6.0-6.8 \\
\cite{2011MNRAS.413.1479S} & 6.6-7.1 \\
\cite{2013ApJ...768...76S} & 5.1-6.6/6.9-7.5\tablefootmark{d} \\
Kormendy \& Ho (2013) & 6.4-7.2 \\

\hline
\end{tabular}
\tablefoot{Note that the derivation of uncertainties for black hole masses is complex since it includes measurement uncertainties (e.g. of the flux measurement) but also the intrinsic scatter of the relations. Usually, the uncertainty will be at least 0.5 dex.~~ \tablefootmark{(a)}{correction for the difference between near-infrared ($\sigma_\mathrm{CO}$) and optical ($\sigma_\mathrm{opt}$) stellar velocity dispersion according to \cite{2014arXiv1410.7723R}.} \tablefoottext{b}{Using the relation for barred galaxies.} \tablefoottext{c}{Note that the $M_\mathrm{BH}-L_\mathrm{bulge}$ relation is not a good estimator for active galaxies according to \cite{2014A&A...561A.140B}.} \tablefoottext{d}{Using the relation for core-S\'ersic galaxies or S\'ersic galaxies respectively.}}
\end{table}

For 74 galaxies from the LLQSO sample, BH masses are available from \cite{2009A&A...507..781S}, \cite{2010A&A...516A..87S}, and Schulze (priv. comm.). From single epoch spectroscopy, they compute BH masses using the scaling relation between broad line region (BLR) size and continuum luminosity \citep{2009ApJ...697..160B}
\begin{equation}
M_\mathrm{BH} = 6.7\, f\, \left( \frac{L_{5100}}{10^{37}\,\mathrm{W}} \right)^{0.52} \left(\frac{\sigma_{\mathrm{H}\beta}}{\mathrm{km}\,\mathrm{s}^{-1}} \right)^2\, M_\odot .
\end{equation}
We use their measurements for $L_{5100}$ and $\sigma_{\mathrm{H}\beta}$ and apply the recent scale factor $f=5.9$ for active galaxies \citep{2013ApJ...772...49W}. The resulting black hole mass for HE 1029-1831 is $\log(M_\mathrm{BH}/M_\odot) = 7.4$.

Scharw\"achter et al. (in prep.) use QDeblend$^\mathrm{3D}$ \citep{2013A&A...549A..43H,2014MNRAS.443..755H} in order to deblend QSO and host galaxy emission in optical IFS-datacubes obtained with WIFES \citep{2007Ap&SS.310..255D,2010Ap&SS.327..245D} and estimate a black hole mass of $\log(M_\mathrm{BH}/M_\odot) = 7.3$ from $L_{5100}$ and the FHWM of the broad H$\beta$ component.

We derive the black hole mass from the broad Pa$\alpha$ component following \cite{2010ApJ...724..386K}
\begin{equation}
M_\mathrm{BH} = 10^{7.16\pm 0.04} \left( \frac{L_{\mathrm{Pa}\alpha}}{10^{35}\,\mathrm{W}} \right)^{0.49\pm 0.06} \left(\frac{\mathrm{FWHM}_{\mathrm{Pa}\alpha}}{10^3\,\mathrm{km}\,\mathrm{s}^{-1}} \right)^2 \,M_\odot.
\end{equation}
We measure a luminosity in the broad Pa$\alpha$ component of $L_{\mathrm{Pa}\alpha,\mathrm{broad}} = 1.04\times 10^{34}\,\mathrm{W}$. With a broad line width of $\mathrm{FWHM}=2150\,\mathrm{km}\,\mathrm{s}^{-1}$, we get a black hole mass of $\log{M_\mathrm{BH}}=7.3$. However, because of the uncertainties discussed in Sect.~\ref{sec:ext}, we did not correct for extinction. Extinction correction would result in a higher estimate.

The stellar velocity dispersion, derived from the stellar continuum fit in Sect.~\ref{sec:stellar}, can be used to estimate the black hole mass from the $M_\mathrm{BH}-\sigma_*$ relation. From the velocity dispersion $\sigma_*=104\,\mathrm{km}\,\mathrm{s}^{-1}$, we find the following estimates for $\log(M_\mathrm{BH}/M_\odot)$: 7.0, 6.7, and 7.4 \citep[using the relations of][]{2009ApJ...698..198G,2011MNRAS.412.2211G,2013ARA&A..51..511K}. Recently, \cite{2014arXiv1410.7723R} reported that the stellar velocity dispersion in spiral galaxies derived from \ion{CO}{} band heads in the near-infrared ($\sigma_\mathrm{CO}$) and from the Calcium triplet in the optical ($\sigma_\mathrm{opt}$) differ from each other significantly. In the above mentioned $M_\mathrm{BH}-\sigma$ relations, the optical velocity dispersion is used. For HE 1029--1831, we correct our near-infrared velocity dispersion using their best fit and derive a corresponding velocity dispersion of $\sigma_\mathrm{opt}=134\,\mathrm{km}\,\mathrm{s}^{-1}$ in the optical. This results in BH masses that are higher by a factor of $\approx 0.5\,\mathrm{dex}$ what is in good consistency with the black hole masses derived with other methods.

One has to note that the BH masses derived from the BH-mass scaling relations like the $M_\mathrm{BH}-\sigma$ relation are only estimations. The uncertainty will be high, at least $0.5\,\mathrm{dex}$, since it includes measurement uncertainties but also the intrinsic scatter of the relations.

With the bulge magnitude $M_\mathrm{bulge}=-23.57$ (Sect.~\ref{sec:decomp}), we can estimate the black hole mass that would be expected from $M_\mathrm{BH}-L_\mathrm{bulge}$ relations for inactive galaxies. The mass estimates $\log(M_\mathrm{BH}/M_\odot)$ range from 8.0 to 8.4 \citep[following][]{2003ApJ...589L..21M,2012MNRAS.419.2264V,2013ApJ...764..151G,2013ARA&A..51..511K}. As observed previously for other LLQSOs, the measured black hole masses are lower by a factor of $\approx 1$ dex than the black hole masses from the $M_\mathrm{BH}-L_\mathrm{bulge}$ relations of inactive galaxies. As discussed in \cite{2014A&A...561A.140B}, this could be caused by overluminous host galaxies that contain young stellar populations or by black holes that are in a growing phase and still undermassive in comparison to their inactive counterparts. 

The Eddington ratio is a measure of the accretion efficiency and is defined as $\lambda \equiv L_\mathrm{bol}/L_\mathrm{Edd}$. The bolometric luminosity can be derived from the X-ray luminosity. The soft X-ray flux from ROSAT has been measured to be $f_\mathrm{soft\,X-ray} = (2.93\pm0.49)\times 10^{-15}\,\mathrm{W}\,\mathrm{m}^{-2}$ \citep{2010MNRAS.401.1151M}, corresponding to a luminosity of $L_\mathrm{soft\,X-ray} = (1.1\pm0.2)\times 10^{36}\,\mathrm{W}$. With a conversion factor of $\approx 50$ \citep{2007ApJ...654..731H}, we get a bolometric luminosity of $L_\mathrm{bol,X-ray}=(6\pm1)\times 10^{37}\,\mathrm{W}$. 

Another possibility is to derive the bolometric luminosity from the AGN continuum luminosity at $5100\,\AA$. Scharw\"achter et al. (in prep.) subtract the host galaxy contribution using QDeblend$^\mathrm{3D}$ and derive a flux of $F_{5100\,\AA}=1.0\times10^{-15}\,\mathrm{erg}\,\mathrm{s}^{-1}\,\mathrm{cm}^{-2}\,\mathrm{\AA}^{-1}$. Using $L_\mathrm{bol}=f_L\times L_{5100\,\AA}$ with $f_L=9$ following \cite{2000ApJ...533..631K}, this results in a bolometric luminosity of $L_\mathrm{bol}=1.9\times 10^{37}\,\mathrm{W}$.

The Eddington luminosity is given by $L_\mathrm{Edd} \cong 1.26 \times 10^{31} \left(M_\mathrm{BH}/M_\odot\right) \,\mathrm{W}$. 
Thus, the range of black hole mass estimates of $\log(M_\mathrm{BH}) = (6.7 - 7.4)$ corresponds to Eddington luminosities $L_\mathrm{Edd}= (6-32)\times 10^{37}\,\mathrm{W}$. With the bolometric luminosities $L_\mathrm{bol} = (2 - 6)\times 10^{37}\,\mathrm{W}$, we can estimate the Eddington ratio to be around $\lambda = (0.06 - 1.0)$, with a mean of $\lambda = 0.3$. Although the range is very broad, we can point out that even the lowest possible Eddington ratio is high, indicating an actively accreting black hole, i.e. a central black hole in a phase of substantial growth.

\subsection{Emission line excitation mechanisms}

In our spectra, we observe a variety of emission lines, most of them originate from ionized or molecular hydrogen but also [\ion{Fe}{ii}]. In recent 3D spectroscopy studies, it has been shown that the H$_2$ and the [\ion{Fe}{ii}] emitting gas often have different flux distributions and kinematics. We find that [\ion{Fe}{ii}] and hydrogen recombination lines (Pa$\alpha$ and Br$\gamma$) show similar distribution while H$_2$ is more centrally distributed (see Figs. \ref{fig:paa} and \ref{fig:h2maps}). The \mbox{AGNIFS} group interprets the distinct flux distributions and kinematics of molecular gas and [\ion{Fe}{ii}] as an indication that molecular gas is rather a tracer of the feeding of the AGN while [\ion{Fe}{ii}] is more likely a tracer of the feedback \citep[e.g.,][]{2006MNRAS.373....2R,2008MNRAS.385.1129R,2011MNRAS.411..469R,2011MNRAS.417.2752R}.

Rotational and vibrational H$_2$ emission lines are in particular the dominant way to cool warm molecular gas. Line ratios of these transitions and ionized hydrogen emission lines give important information on their excitation mechanisms \citep{2003ApJ...597..907D,2005ApJ...633..105D}. In the case of HE 1029-1831, they indicate a significant contribution of currently occuring star formation and/or young stellar populations in the central region. 

In general, the H$_2$ emission lines can be excited by two kinds of processes: (1) thermal processes, e.g. produced by X-ray \citep{1996ApJ...466..561M} or by shock heating \citep{1989ApJ...342..306H} and (2) non-thermal processes like UV fluorescence \citep{1987ApJ...322..412B}: UV-photons with $912\AA<\lambda<1500\AA$ are absorbed by the H$_2$ molecule in the Lyman- and Werner bands, exciting the next two electronic levels ($B^1\Sigma-X^1\Sigma$ and $C^1\Pi-X^1\Sigma$). With a probability of 90\%, a decay into a bound but excited rovibrational level within the electronic ground level $X^1\Sigma$ will take place. By this mechanism, H$_2$ rovibrational levels will be populated, which could not be populated by collisions. Possible sources are OB stars or strong AGN continuum emission.

Different excitation mechanisms can be discriminated by observing H$_2$ line ratios. In particular, the H$_2$ line ratio 2-1S(1)/1-0S(1) is useful to distinguish between thermal and non-thermal processes while H$_2$ 1-0S(2)/1-0S(0) can be used to distinguish between excitation by X-rays and by shocks. Figure \ref{fig:h2excitation} (left) shows the diagnostic diagram with the H$_2$ line ratios 2-1S(1)/1-0S(1) vs. 1-0S(2)/1-0S(0) with the two chosen apertures in blue and red \citep[see e.g.,][]{1994ApJ...427..777M,2007A&A...466..451Z,2013MNRAS.428.2389M,2014A&A...567A.119S,2014MNRAS.438..329F}. Furthermore, we indicate the positions of models for thermal UV excitation \citep{1989ApJ...338..197S}, UV fluorescence \citep{1987ApJ...322..412B}, X-ray heating \citep{1990ApJ...363..464D}, shock-heating \citep{1989MNRAS.236..929B}, and a thermal emission curve. The position in the diagram shows that the H$_2$ excitation in the two apertures cannot be explained by thermal excitation alone. A significant contribution from UV fluorescence is necessary. This is in good consistency with star formation activity in the two spots. {Previous studies show that in the case of AGN, collisional excitation, e.g. by interaction of the radio-jet with the circumnuclear interstellar medium, might be more important than the fluorescent process \citep{2006MNRAS.373....2R,2010MNRAS.404..166R}. However, other studies see strong evidence that fluorescent excitation plays indeed a major role in the H$_2$ excitation in ULIRGs \citep{2003ApJ...597..907D} and AGNs \citep{2005ApJ...633..105D}, in concordance with our findings.

The rotational temperature can be determined from two ortho/para lines that belong to the same vibrational level, whereas the vibrational temperature can be determined by connecting two transitions with same $J$ but from consecutive $v$ levels:
\begin{eqnarray}
T_{\mathrm{rot}(v=1)} = \frac{1113\,\mathrm{K}}{1.130+\ln \left( \frac{f_\mathrm{1-0\, S(0)}}{f_\mathrm{1-0\, S(2)}} \right)} 
\label{eq:temprot}\\
T_\mathrm{vib} = \frac{5594\,\mathrm{K}}{0.304+\ln \left( \frac{f_\mathrm{1-0\, S(1)}}{f_\mathrm{2-1\, S(1)}} \right)} .
\label{eq:tempvib}
\end{eqnarray}
Using the flux measurements listed in Tab.~\ref{tab:emline-fluxes}, we estimate a rotational temperature of $T_\mathrm{rot}=(1100\pm300)\,\mathrm{K}$ in spot 1 and $T_\mathrm{rot}=(900\pm200)\,\mathrm{K}$ in spot 2. The vibrational temperature is $T_\mathrm{vib}(\mathrm{spot 1})=(3700\pm600)\,\mathrm{K}$ and $T_\mathrm{vib}(\mathrm{spot 2})=(3200\pm500)\,\mathrm{K}$. In Fig.~\ref{fig:h2excitation} (left), the rotational and vibrational temperatures are shown as right and upper axis. 

In case of purely thermal emission, we expect the H$_2$ molecules to be in local thermal equilibrium. In this case, $T_\mathrm{vib}\approx T_\mathrm{rot}$. In particular, \cite{2005MNRAS.364.1041R} and \cite{2013MNRAS.430.2002R} show that star forming galaxies show $T_\mathrm{vib}\gtrsim T_\mathrm{rot}$ while AGNs and LINERs tend to have similar values for $T_\mathrm{vib}$ and $T_\mathrm{rot}$. In HE 1029-1831, the temperatures differ significantly from each other. This points at a significant contribution from the non-thermal process of UV-fluorescence, consistent with star formation in the circumnuclear region.

\begin{figure*}
\includegraphics[width=0.48\linewidth]{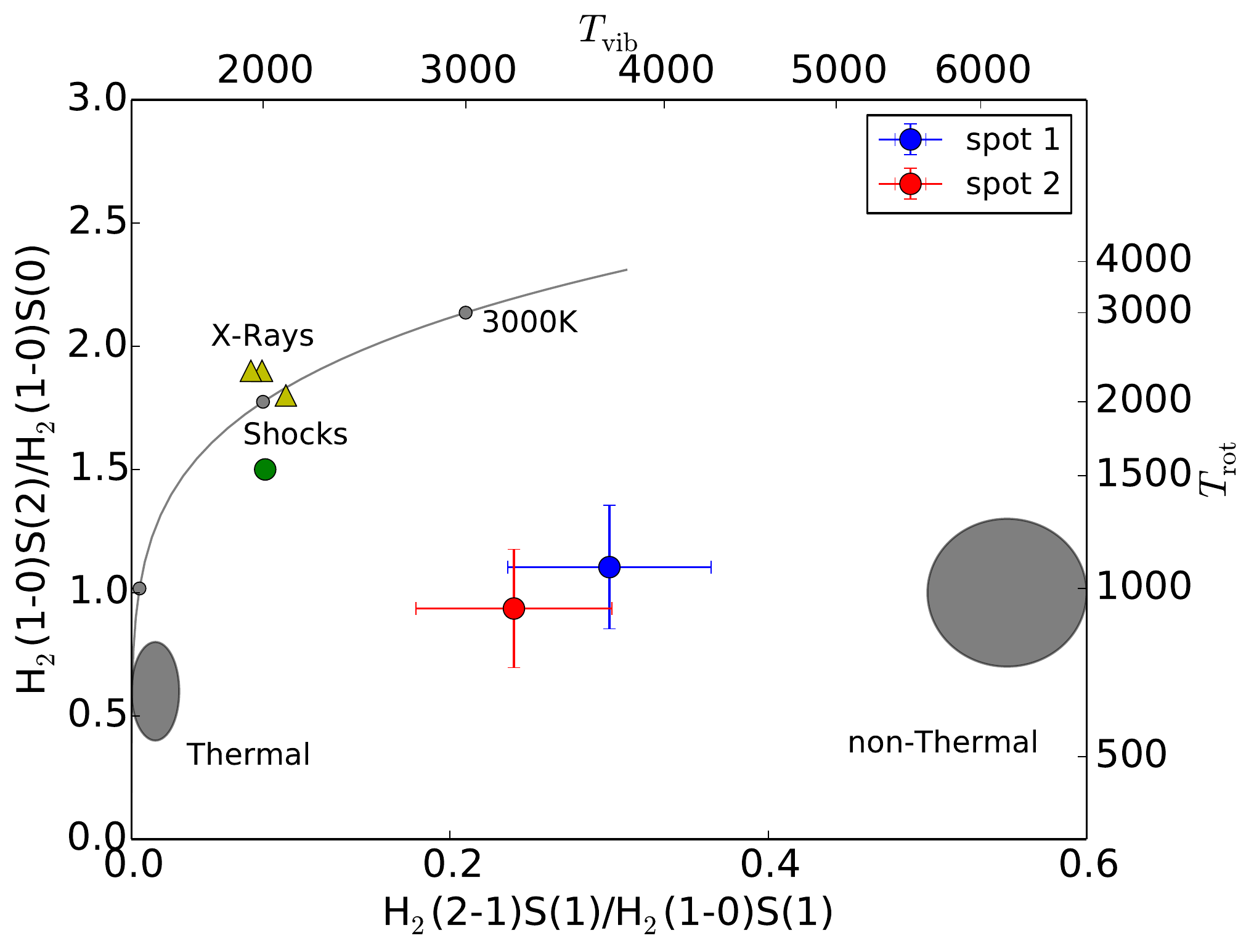}
\includegraphics[width=0.48\linewidth]{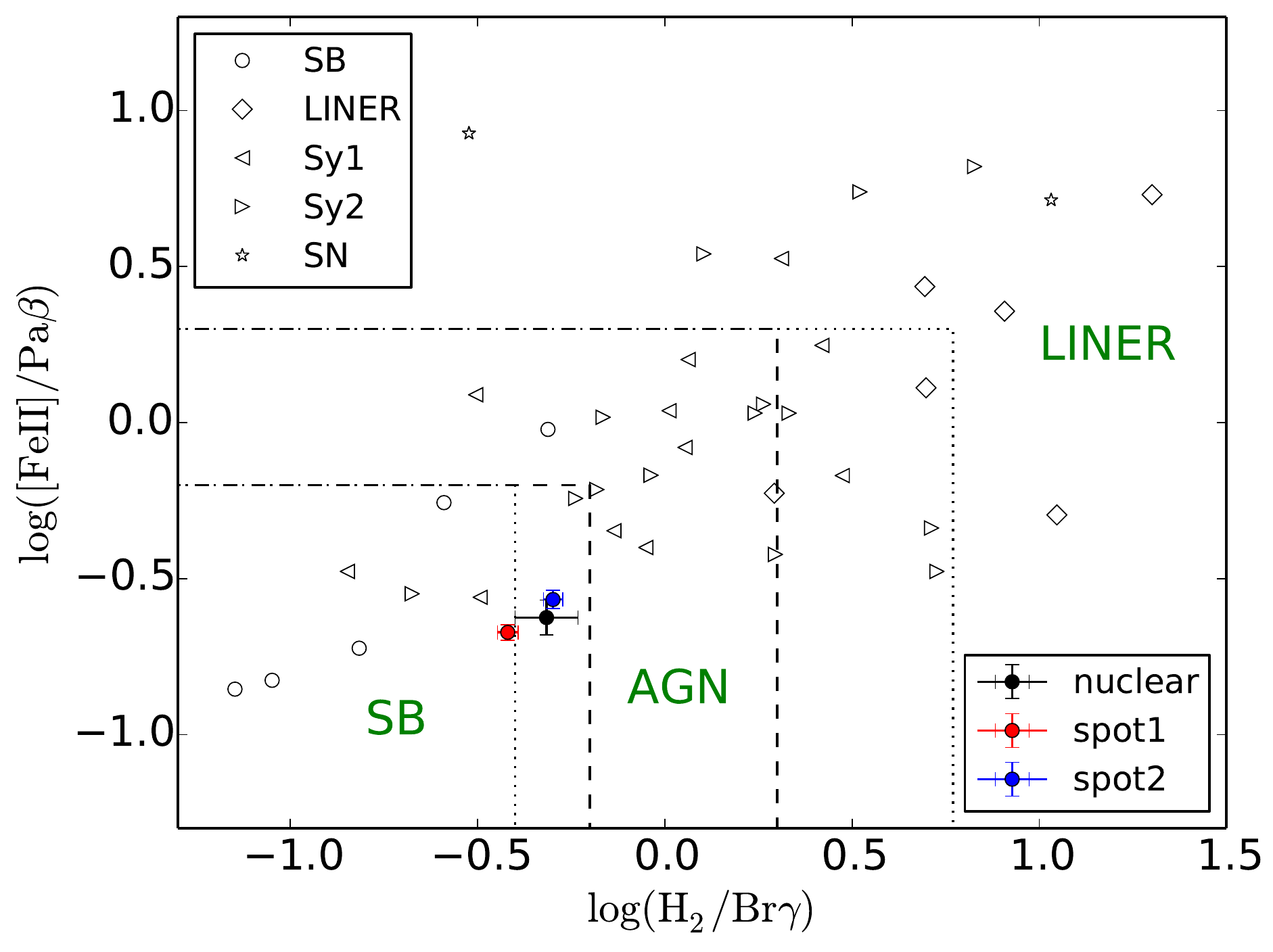}
\caption{\emph{Left:} Molecular hydrogen diagnostic diagram with 2-1S(1)/1-0S(1) vs. 1-0S(3)/1-0S(1) \citep{1994ApJ...427..777M}. Spot 1 and 2 are indicated by blue and red points. The location of thermal UV excitation \citep{1989ApJ...338..197S} and non-thermal models \citep{1987ApJ...322..412B}, as well as the thermal emission curve from 1000 K to 3000 K are plotted in gray. The X-ray heating models \citep{1990ApJ...363..464D} are marked by a yellow triangle, the shock-heating model \citep{1989MNRAS.236..929B} by a green circle. \emph{Right:} Diagnostic diagram with line ratios 1-0S(1)$\lambda 2.121\mu$m/Br$\gamma$ vs. [\ion{Fe}{ii}]$\lambda1.257\mu$m/Pa$\beta$ (for details see text). The positions of three apertures are indicated by circles. Open symbols correspond to literature values from \cite{1998ApJS..114...59L,2004ApJ...601..813D,2004A&A...425..457R,2005MNRAS.364.1041R}. The lines indicate regions that are typically populated by starburst galaxies, AGNs, and LINERs resp. Dashed lines: \cite{2005MNRAS.364.1041R,2010MNRAS.404..166R}, dotted lines: \cite{2013MNRAS.430.2002R}.}
\label{fig:h2excitation}
\end{figure*}

Further support for this hypothesis is given by the line ratios of $\log ([\ion{Fe}{ii}]/\mathrm{Br}\gamma)$ that are expected to be $<0.3$ for star formation regions \citep{1997ApJ...482..747A}. In our three apertures we find $\log ([\ion{Fe}{ii}]/\mathrm{Br}\gamma)_\mathrm{nucl}=0.04$, $\log ([\ion{Fe}{ii}]/\mathrm{Br}\gamma)_\mathrm{spot1}=-0.01$, and $\log ([\ion{Fe}{ii}]/\mathrm{Br}\gamma)_\mathrm{spot2}=0.05$.

\cite{2004A&A...425..457R,2005MNRAS.364.1041R} suggest a NIR diagnostic diagram with the line ratios $\log(\mathrm{H}_2(1-0)\mathrm{S}(1)/\mathrm{Br}\gamma)$ and $\log ([\ion{Fe}{ii}]\lambda 1.257\mu\mathrm{m}/\mathrm{Pa}\beta)$. The diagnostic is supposed to distinguish between excitation from pure photoionization and from pure shocks and shows a transition from starburst galaxies to LINERS, passing AGNs where both excitation mechanisms are of importance. Since we are lacking $J$-band information, we use the conversion factors $[\ion{Fe}{ii}]\lambda 1.644\mu\mathrm{m}/[\ion{Fe}{ii}]\lambda 1.257\mu\mathrm{m}=0.744$ \citep{1988A&A...193..327N} and $\mathrm{Pa}\alpha/\mathrm{Pa}\beta=2.05$ \citep{2006agna.book.....O}. The line ratios $[\log(\mathrm{H}_2/\mathrm{Br}\gamma),\log ([\ion{Fe}{ii}]/\mathrm{Pa}\beta)]=[-0.32,-0.60]$ for the nuclear region, $[-0.42,-0.66]$ for spot1, and $[-0.30,-0.59]$ for spot2 are consistent with a mixture of starburst activity and AGN excitation (see Fig.~\ref{fig:h2excitation} right). Previous studies \citep[e.g.,][]{2010MNRAS.404..166R,2014MNRAS.445..414S} show that the centers of active galaxies often show line ratios characteristic for a mixture of starburst and AGN excitation.

\subsection{Mass of ionized and molecular gas}
\label{sec:gasmasses}

A sufficient reservoir of gas is necessary to fuel star formation and black hole accretion. We show that the LLQSO HE 1029-1831 is rich in ionized and molecular gas.

We estimate the mass of ionized hydrogen as $M_\ion{H}{ii}=m_\mathrm{p} n_\mathrm{e} V_\ion{H}{ii}$ with the electron density $n_\mathrm{e}$ and the volume $V_\ion{H}{ii}$ of the emitting region. Using the line coefficients from \cite{2006agna.book.....O}, assuming an electron temperature of $T=10^4\,\mathrm{K}$ and a density in the range $10^2 < n_\mathrm{e} < 10^4\,\mathrm{cm}^{-3}$, we can calculate the Pa$\alpha$ flux as
\begin{eqnarray}
f_{\mathrm{Pa}\alpha} &=& \frac{\iint j_{\mathrm{Pa}\alpha}\,\mathrm{d}\Omega\,\mathrm{d}V}{4\pi d^2} 
= \frac{1}{4\pi} \left( \frac{4\pi j_{\mathrm{H}\beta}}{n_\mathrm{e}^2} \right) \left(\frac{j_{\mathrm{Pa}\alpha}}{j_{\mathrm{H}\beta}} \right) \frac{n_\mathrm{e}^2 V_\ion{H}{ii}}{D^2} \\
&\approx& 3.3\times 10^{-27}\, \frac{n_\mathrm{e}^2 V_\ion{H}{ii}}{D^2}\,\mathrm{erg}\,\mathrm{cm}^{-2}\,\mathrm{s}^{-1}
\end{eqnarray}
where $D$ is the distance to the galaxy in cm. The ionized gas mass is then given by
\begin{equation}
M_\ion{H}{ii} \approx 2.1\times 10^{21}\, \left(\frac{f_{\mathrm{Pa}\alpha}}{\mathrm{W}\,\mathrm{m}^{-2}} \right) \left(\frac{D}{\mathrm{Mpc}}\right)^2 \left(\frac{n_\mathrm{e}}{\mathrm{cm}^{-3}} \right)^{-1} \,M_\odot
\end{equation}
In an aperture with radius $4 \arcsec$, the flux in Pa$\alpha$ is $f_{\mathrm{Pa}\alpha}(r\leq 4\arcsec)=43\times 10^{-18}\,\mathrm{W}\,\mathrm{m}^{-2}$. Assuming an electron density of $n_e=100\,\mathrm{cm}^{-3}$, this corresponds to a mass in ionized hydrogen of $M_\ion{H}{ii}(r\leq 4\arcsec)=2.9\times 10^7\,M_\odot$.

The mass of warm H$_2$ can be estimated by 
\begin{equation}
M_{\mathrm{H}_2} \approx 5.0776 \times 10^{16} \left( \frac{F_{\mathrm{H}_2}}{\mathrm{W}\,\mathrm{m}^{-2}} \right) \left( \frac{D}{\mathrm{Mpc}} \right)^2 M_\odot
\end{equation}
following \cite{1982ApJ...253..136S}, \cite{1998ApJS..115..293W} and \cite{2010MNRAS.404..166R}. In an aperture of radius $4\arcsec$, we measure a H$_2$(1-0)S(1) flux $f_{\mathrm{H}_2}=290\times  10^{-20}\,\mathrm{W}\,\mathrm{m}^{-2}$. This corresponds to a warm H$_2$ gas mass of $4700\,M_\odot$. We see that the mass of ionized gas is a factor of $\approx 6000$ higher than the mass of warm molecular hydrogen. This is in good agreement with typical ratios that are of the order $10^3 - 10^4$ \citep[][and references therein]{2014MNRAS.442..656R}. With the conversion factor of \cite{2013MNRAS.428.2389M}, $M_{\mathrm{H}_2\mathrm{(cold)}}/M_{\mathrm{H}_2\mathrm{(warm)}} = (0.3 - 1.6) \times 10^6$, we derive a cold gas mass of $(1.4 - 7.5)\times 10^9\,M_\odot$. Other cold-to-warm $H_2$ gas mass ratios have been found by \cite{2005AJ....129.2197D} ($10^5-10^7$) or \cite{2006A&A...454..481M} ($2.5\times 10^6$ with a 1$\sigma$ uncertainty of a factor of 2) which result in cold gas masses that are consistent with our results.

\cite{2007A&A...464..187K} and Moser et al. (in prep.) derive the molecular gas mass from CO(1-0) measurements. They estimate masses of a few to several $10^9\,M_\odot$, which is in good consistency with our results.

For NUGA sources, cold molecular gas masses have been calculated from CO-emission, ranging from $2\times 10^8- 2\times 10^{10}\,M_\odot$ with typical masses of the order of several $10^9\,M_\odot$ \citep[][and references therein]{2012nsgq.confE..69M}. The \mbox{AGNIFS} group obtained masses in a range of $66\,M_\odot \leq M_{\mathrm{H}_2} \leq 3300\,M_\odot$ and $0.1\times 10^6\,M_\odot \leq M_{\ion{H}{ii}} \leq 17\times 10^6\,M_\odot$ in the nuclear regions of nearby galaxies \citep{2008MNRAS.385.1129R,2009MNRAS.393..783R,2009MNRAS.394.1148S,2010MNRAS.404..166R,2014MNRAS.445..414S}. We conclude that the LLQSO HE 1029-1831 has a large reservoir of cold molecular gas that is needed and apparently used for both, star formation and black hole fueling.

\subsection{The impact of star formation in the bulge}
\label{sec:sf}

In our previous study \citep{2014A&A...561A.140B}, we found that the observed LLQSOs do not follow published $M_\mathrm{BH}-L_\mathrm{bulge}$ relations for inactive galaxies and speculated about the influence of young or intermediate-age stellar populations in the bulge lowering the mass-to-light ratio. Here, we estimate star formation rates and the dynamical mass of the bulge. Furthermore, we use the equivalent width of Br$\gamma$, the supernova rate, and the mass-to-light ratio as diagnostics to contrain possible star formation histories and estimate the impact of star formation in the bulge.

\subsubsection{Star formation rates}

Since dust grains absorb the radiation of young stars and re-radiate it in the far-infrared (FIR), the FIR luminosity $L_\mathrm{FIR}$ is a tracer of star formation activity on 100-300 Myr timescales.

From the IRAS faint source catalog \citep{1992ifss.book.....M}, we calculate the FIR luminosity $L_\mathrm{FIR} = 4\pi \left(D_L/\mathrm{Mpc}\right)^2 \times 1.2\times 10^{31} \left(2.58\,f_{60} + f_{100}\right)\,\mathrm{W} = 4.9 \times 10^{37}\,\mathrm{W}$, where $f_\lambda$ is the flux density at $60\,\mu\mathrm{m}$ and $100\,\mu\mathrm{m}$ resp. \citep{1988ApJS...68..151H,1996ARA&A..34..749S}. With the calibration of \cite{2003A&A...409...99P}, we estimate a global star formation rate of
\begin{equation}
\mathrm{SFR}_\mathrm{FIR} = \frac{L_\mathrm{FIR}}{1.134 \times 10^{36}\,\mathrm{W}} = 42.9\,M_\odot\,\mathrm{yr}^{-1}
\end{equation}
This FIR luminosity could be heavily affected by the AGN emission. Therefore, the star formation rate should only be taken as an upper limit.

Furthermore, we calculate the FIR colors $\log(f_{25}/f_{12})=0.47\pm0.13$, $\log(f_{60}/f_{25})=0.79\pm0.05$, $\log(f_{100}/f_{25})=0.95\pm0.12$, and $\log(f_{60}/f_{12})=1.26\pm0.12$. According to \cite{1998ApJ...498..570D,2000ApJ...530..704K}, these ratios are in the range of AGN with more than 90\% contribution to the FIR luminosity from star formation. Together with $\mathrm{SFR}_\mathrm{FIR}$ calculated above, this implies that HE 1029-1831 is actively star forming.

With our high-resolution data, we can calculate current star formation rates in the spatially resolved nuclear regions. We use Br$\gamma$ as a star formation indicator according to \cite{2003A&A...409...99P}
\begin{equation}
\frac{\mathrm{SFR}}{M_\odot\,\mathrm{yr}^{-1}} = \frac{L(\mathrm{Br}\gamma)}{1.585\times 10^{32}\,\mathrm{W}}.
\end{equation}
In the two off-nuclear spots, we measure $\mathrm{SFR}_\mathrm{spot1}=0.89\,M_\odot\,\mathrm{yr}^{-1}$ and $\mathrm{SFR}_\mathrm{spot2}=0.52\,M_\odot\,\mathrm{yr}^{-1}$. This corresponds to star formation rate densities of $\Sigma_\mathrm{SFR,spot1}=19\,M_\odot\,\mathrm{yr}^{-1}\,\mathrm{kpc}^{-2}$ and $\Sigma_\mathrm{SFR,spot2}=12\,M_\odot\,\mathrm{yr}^{-1}\,\mathrm{kpc}^{-2}$. Our aperture has a radius of $0\farcs15 = 120\,\mathrm{pc}$. Thus, our values are in agreement with the expected star formation rate densities in Seyfert galaxies of $(1-50)\,M_\odot\,\mathrm{yr}^{-1}\,\mathrm{kpc}^{-2}$ on hundreds of parsecs scales and $(50-500)\,M_\odot\,\mathrm{yr}^{-1}\,\mathrm{kpc}^{-2}$ on tens of parsecs scales \citep[][and references therein]{2012A&A...544A.129V}.

Furthermore, we can compare the star formation rate densities to the gas mass densities (estimated from the hot molecular gas mass) and find $\log \left( \Sigma_\mathrm{gas,\mathrm{spot1}}/M_\odot\,\mathrm{pc}^{-2} \right)=3.7$ and $\log \left( \Sigma_\mathrm{gas,\mathrm{spot2}}/M_\odot\,\mathrm{pc}^{-2} \right)=3.6$. This means that the star formation regions follow the Schmidt law \citep{1998ApJ...498..541K} and the high gas mass is efficiently transformed into stars.

\subsubsection{Dynamical bulge mass}

The combination of the image analysis that provides us with the effective radius and the kinematical analysis that provides us with the stellar velocity dispersion, allows us to calculate the dynamical mass of the central bulge component. \cite{2011MNRAS.413.1479S}, following \cite{2006MNRAS.366.1126C}, adopt
\begin{equation}
M_\mathrm{dyn} = \frac{\kappa r_e \sigma^2}{G}
\end{equation}
with $\kappa=5$ to derive the dynamical mass in a bulge with effective radius $r_e$. With the values for $r_e$ and $\sigma_e$ derived in Sections \ref{sec:decomp} and \ref{sec:stellar}, this results in $M_\mathrm{dyn}=6\times 10^9\,M_\odot$.

From the H$_2$ flux in an aperture with radius $r_e$, we can estimate the cold H$_2$ gas mass in the bulge region (see Sect.~\ref{sec:gasmasses}). We multiply with 1.36 in order to allow for the contribution of Helium gas and obtain $(1.0-4.9)\times 10^9\,M_\odot$. This corresponds to a gas fraction of $20\%-80\%$, depending on the warm-to-cold gas conversion factor applied. We subtract the gas mass from the derived dynamical mass to get an estimate for the stellar mass only: $M_* = (1.1-5.0)\times 10^9\,M_\odot$.

Several correlations between the black hole mass and the (stellar) bulge mass have been found in the last years. We use the dynamical bulge mass estimate to find the corresponding black hole mass: $\log (M_\mathrm{BH}/M_\odot) = 6.0 - 7.5$ \citep[using the relations of][]{2004ApJ...604L..89H,2011MNRAS.413.1479S,2013ApJ...768...76S,2013ARA&A..51..511K}. Interestingly, these estimates agree well with the black hole masses derived in Sect. \ref{sec:bh}, in contrast to estimates from $M_\mathrm{BH}-L_\mathrm{bulge}$ relations. This can be interpreted as hint that the deviation of LLQSOs from the $M_\mathrm{BH}-L_\mathrm{bulge}$ relation observed in \cite{2014A&A...561A.140B} is - at least in the case of HE 1029-1831 - primarily caused by the higher luminosity of young stellar populations rather than a significantly undermassive black hole.

\subsubsection{Star formation history}

To follow this trace, we use \textsc{Starburst99} \citep{1999ApJS..123....3L,2005ApJ...621..695V,2010ApJS..189..309L,2014ApJS..212...14L} to model the mass-to-light ratio ($M_*/L_K$ ratio), Br$\gamma$ equivalent width, and supernova rate which can then be compared to observations \citep{2006ApJ...646..754D,2007ApJ...671.1388D,2013ApJ...779..170L}. We show that the bulge is dominated by young/intermediate-age populations ($\sim 100\,\mathrm{Myr}$) and find traces for a declining starburst $\sim 200\,\mathrm{Myr}$ ago.

For the simulations, we use the Padova tracks with AGB stars at solar metallicity and a Kroupa IMF. We consider five different star formation histories: an instantaneous starburst, single starbursts with decay times of $\tau_\mathrm{SF}=10\,\mathrm{Myr}$, $50\,\mathrm{Myr}$, and $100\,\mathrm{Myr}$ as well as continuous star formation. The decaying starbursts were simulated by adding up instantaneous starbursts scaled to exponentially decaying intensity with the respective time scales.

\begin{figure*}
\centering
\includegraphics[width=0.48\linewidth]{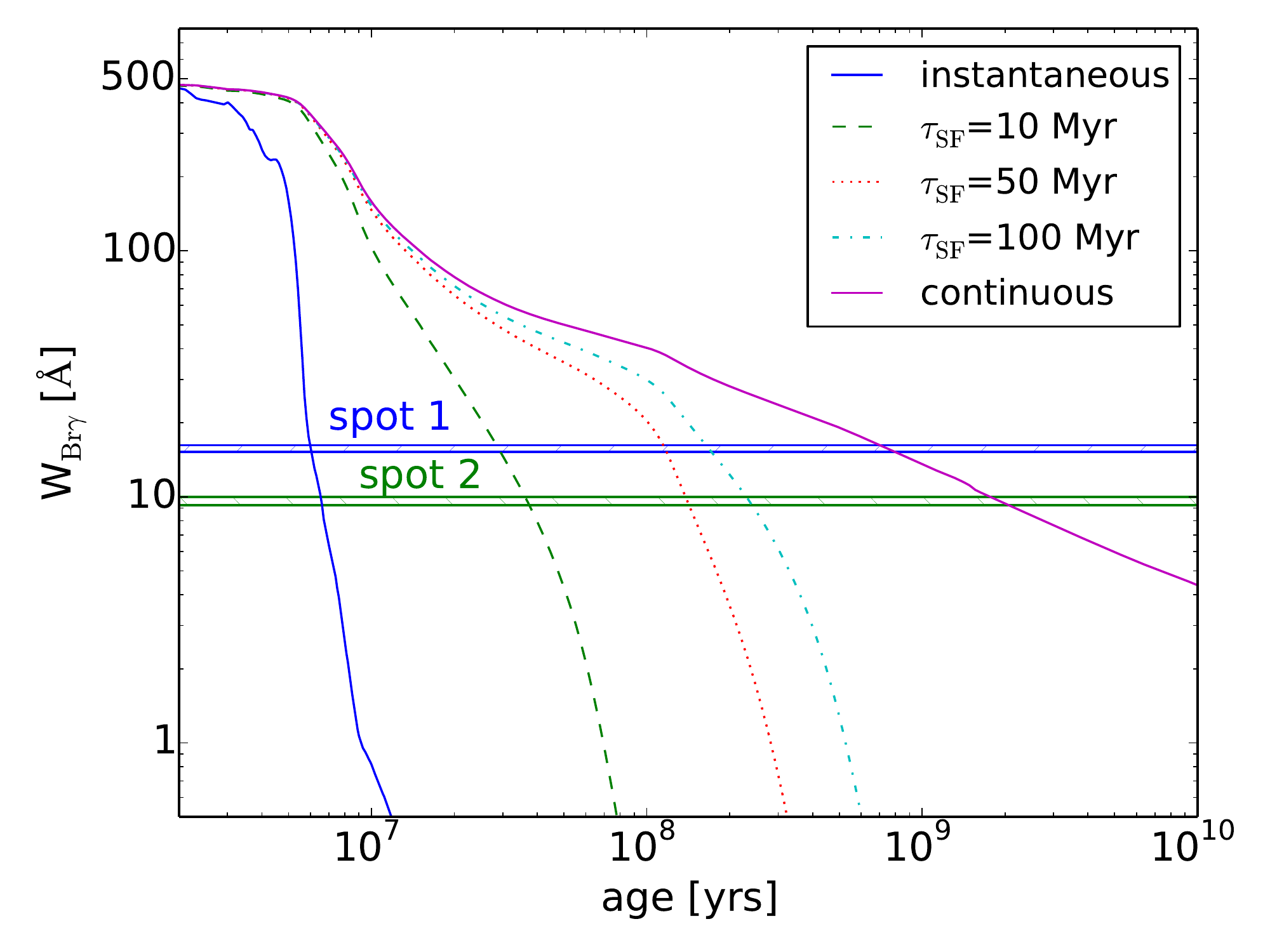}
\includegraphics[width=0.48\linewidth]{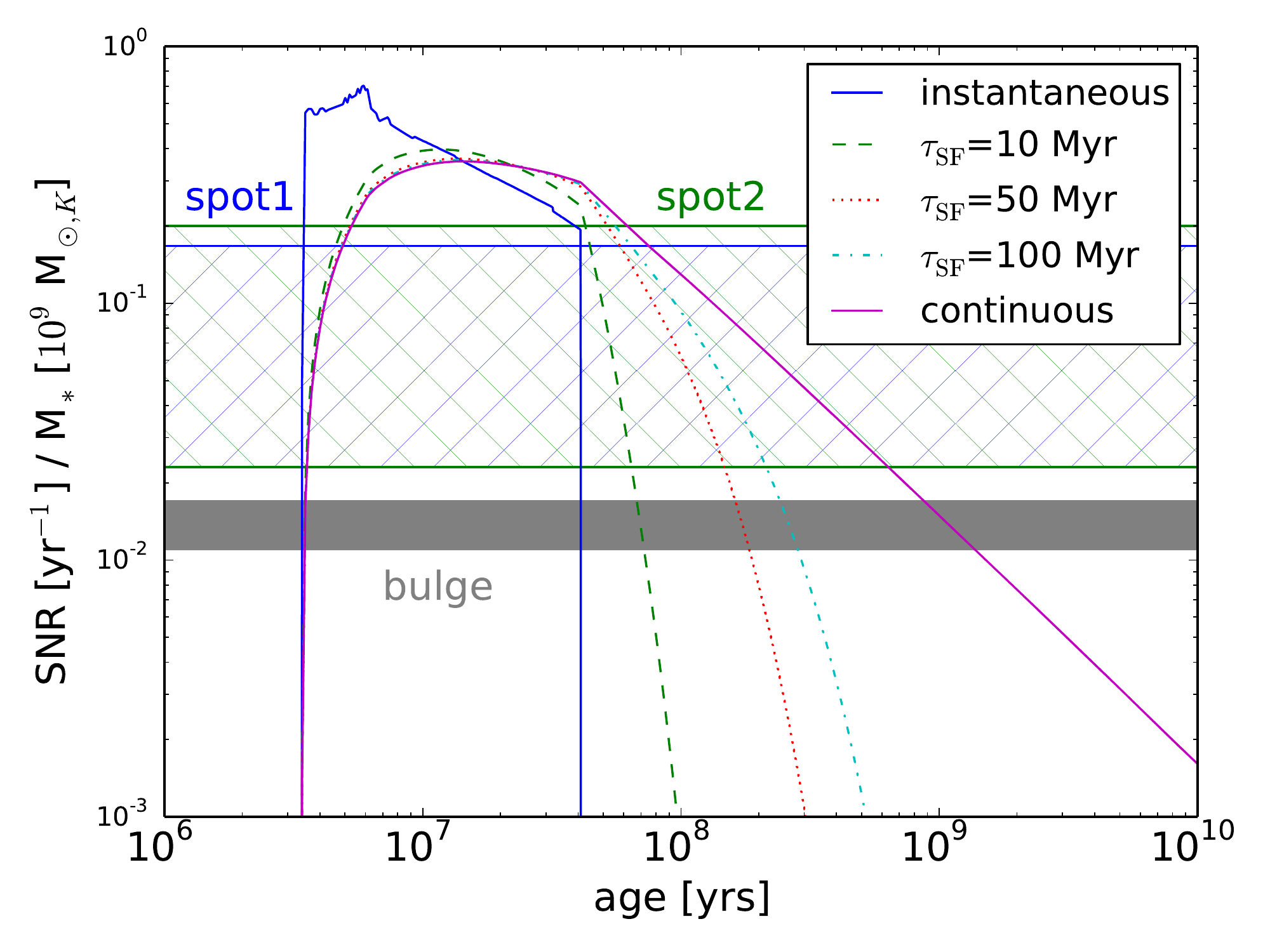}
\caption{Results from the \textsc{Starburst99} simulations. Br$\gamma$ equivalent width (\emph{left}) and supernova rate (\emph{right}) as a function of age for five star formation histories: Instantaneous starburst (``fixed mass'' in \textsc{Starburst99}), continuous star formation, and three exponentially decaying starbursts. The values derived from our data are shown in gray, blue and green.}
\label{fig:sb99}
\end{figure*}

\paragraph{Br$\gamma$ equivalent width} The equivalent width of the stellar Br$\gamma$ emission is supposed to be a good estimator for the stellar age since it decreases monotonically with time (see Fig. \ref{fig:sb99}, left). However, the measurement could be affected by the nonstellar AGN continuum. \cite{2007ApJ...671.1388D} conclude that every stellar population that contains late-type stars should show an equivalent width of the stellar absorption feature CO(2-0) at $\lambda 2.29\,\mu\mathrm{m}$ of $W_\mathrm{CO(2-0)}\approx 12\,\AA$ independent of the star formation history and age with an uncertainty of about 20\%. Thus, they suggest that the fraction of the stellar light, contaminated by nonstellar-AGN continuum, can be estimated by $f_\mathrm{stellar} = W_\mathrm{obs}/W_\mathrm{intr}$ with the observed equivalent width of CO(2-0), $W_\mathrm{obs}$, and the intrinsic $W_\mathrm{intr} =12\,\AA$. 

We use the definitions of \cite{1993A&A...280..536O} and measure the absorption in the interval $2.2924\,\mu\mathrm{m} - 2.2977\,\mu\mathrm{m}$ and the continuum between $2.2880\,\mu\mathrm{m} - 2.2920\,\mu\mathrm{m}$. This results in equivalent widths of CO(2-0) of $W_\mathrm{CO,nucl}=2.0\,\AA$, $W_\mathrm{CO,spot1}=7.1\,\AA$, and $W_\mathrm{CO,spot2}=9.6\,\AA$, corresponding to stellar-light fractions of $f_\mathrm{stellar,nucl}=17\%$, $f_\mathrm{stellar,spot1}=59\%$, and $f_\mathrm{stellar,spot2}=80\%$. We measure the Br$\gamma$ equivalent width in the three apertures. After correcting for the non-stellar contribution, we gain $W_{\mathrm{Br}\gamma\mathrm{,nucl,corr}}= (8.8\pm1.2)\,\AA$, $W_{\mathrm{Br}\gamma\mathrm{,spot1,corr}}= (15.8\pm0.5)\,\AA$, and $W_{\mathrm{Br}\gamma\mathrm{,spot2,corr}}= (9.6\pm0.4)\,\AA$.

\paragraph{Supernova rate} The forbidden iron transition $[\ion{Fe}{ii}]\,\lambda1.644\mu\mathrm{m}$ is a good shock tracer. Since we see no indication for jets and outflows, we assume that the line is mainly excited by shocks from supernovae. The supernova rate can then be calculated 
\begin{equation}
\mathrm{SNR}_\mathrm{Cal97} = 5.38 \, \frac{L_{[\ion{Fe}{ii}]}}{10^{35}\,\mathrm{W}}\,\mathrm{yr}^{-1}
\end{equation}
following \cite{1997AJ....113..162C} or, following \cite{2003AJ....125.1210A},
\begin{equation}
\mathrm{SNR}_\mathrm{AH03} = 8.08 \, \frac{L_{[\ion{Fe}{ii}]}}{10^{35}\,\mathrm{W}}\,\mathrm{yr}^{-1}.
\end{equation}
We measure the [\ion{Fe}{ii}] emission in an aperture corresponding to the half-light radius $r_e=0\farcs59$ and get $L_{[\ion{Fe}{ii}]}=1.03\times 10^{33}\,\mathrm{W}$. This corresponds to supernova rates of $\mathrm{SNR}_\mathrm{Cal97}=0.055\,\mathrm{yr}^{-1}$ and $\mathrm{SNR}_\mathrm{AH03}=0.083\,\mathrm{yr}^{-1}$. In Fig. \ref{fig:sb99} (right), we show the supernova rate divided by the stellar mass (normalized by $10^{9}\,M_{\odot}$) as simulated by \textsc{Starburst99}. As bulge mass estimate, we use the dynamical mass derived above. 

To estimate the stellar mass in the two spots, we assume that the bulge mass is smoothly distributed following a S\'ersic law with the parameters derived in the bulge-disk decomposition. The mass fraction of the spots can then easily be estimated whereas we determine the supernova rate from the [\ion{Fe}{ii}] flux in the apertures. The supernova rates in the two spots are $\mathrm{SNR} (\mathrm{spot1}) = 0.007 - 0.010\,\mathrm{yr}^{-1}$ and $\mathrm{SNR} (\mathrm{spot2}) = 0.005 - 0.008\,\mathrm{yr}^{-1}$. The supernova rates should be taken as upper limits only, since contamination from the AGN cannot be excluded.

\paragraph{Mass-to-light ratio} Figure \ref{fig:sb99_ml} shows that the ratio of the stellar mass to the $K$-band luminosity can be used as a diagnostic for the age of the stellar population for ages $\geq 10^7\,\mathrm{yr}^{-1}$ since it increases monotonically with time. However, two caveats have to be kept in mind: First, we cannot measure the stellar mass directly. Instead, we will take the dynamical mass corrected for the cold gas mass as an estimate. Second, the model assumes that only one stellar component is discussed. However, we expect to have a mixture of an old population ($\approx 10^{10}\,\mathrm{yr}$) that could be mixed with one or more young populations that have been formed in later starbursts. The age estimate from the mass-to-light ratio can thus only be an upper limit for the young population. With the bulge mass $M_\mathrm{bulge}\approx (1-5)\times 10^9\,M_\odot$ and the bulge magnitude $M_\mathrm{bulge}=-23.57$ derived from the bulge-disk decomposition in Sect. \ref{sec:decomp}, we derive a mass-to-light ratio of $M_*/L_K=0.03 - 0.13$.

\paragraph{}The calculated ratios are presented in Figs.~\ref{fig:sb99} and \ref{fig:sb99_ml} together with the \textsc{Starburst99} models. From these diagrams, immediate conclusions can be drawn: From the equivalent width of Br$\gamma$ and the supernova rate, we can rule out an instantaneous starburst scenario for spots 1 and 2 since from $W_{\mathrm{Br}\gamma}$, we would estimate a starburst age of about 5-6 Myr but in this case the supernova rate should be much higher than observed. Continuous star formation is also rather inconsistent with our models: From $W_{\mathrm{Br}\gamma}$, we expect an age of around 800 Myr, while from the supernova rate the age would be around 200-500 Myr. The favored model is an exponentially decaying starburst with time scale $\tau_\mathrm{SF}=50-100\,\mathrm{Myr}$ that began 100-200 Myr ago. 

For the overall bulge, we can again exclude an instantaneous starburst model from the supernova rates. The favored model for the two spots is consistent with supernova rates and the mass-to-light ratio for the overall bulge, indicating that the bulge is dominated by the star formation regions. However, continuous star formation that began around 500 Myr ago is also consistent with the data.

We conclude that we observe an intermediate-age stellar population with an age of around 100 Myr in the two spots that dominate the overall bulge emission. The star formation time scale is of the order of 50-100 Myr. This implies that the star formation is declining, in accordance with the derived current star formation rates that were high compared to quiescent galaxies but not as extraordinary as expected for a starburst galaxy, and with the high value derived from the FIR luminosity that is sensitive to star formation on timescales of 100-300 Myr. Furthermore, this is in good consistency with the finding of \cite{2007ApJ...671.1388D} that AGNs accreting at lower efficiency ($\lambda \leq 0.1$) have younger starbursts ($\leq 50-100\,\mathrm{Myr}$) while AGNs accreting at higher efficiency ($\lambda\geq 0.1$) have older starbursts ($\geq 50-100\,\mathrm{Myr}$).

To conclude, the three diagnostics hint at star formation as an important factor in the central region of the LLQSO HE 1029-1831. Although the star formation activity has already reached its maximum more than 100 Myr ago, the effect is still visible: The presence of intermediate-age stellar population lowers the mass-to-light ratio by a factor of 10 compared to a bulge purely consisting of an old stellar population (see Fig.~\ref{fig:sb99_ml}). 

Interestingly, the dynamical mass estimate that was derived here is consistent with predictions from the $M_\mathrm{BH}-M_\mathrm{bulge}$ relations. Thus, at least in the particular case of HE 1029-1831, we conclude that the observed offset of LLQSOs from the $M_\mathrm{BH}-L_\mathrm{bulge}$ relations for inactive galaxies is caused by recent star formation activity in the central region. 

An alternative explanation is the presence of a pseudo-bulge which is supported by the low S\'ersic index. Pseudo-bulges are thought not to evolve from galaxy mergers but from secular evolution, therefore not obeying the relations of classical bulges and ellipticals \citep[][and references therein]{2004ARA&A..42..603K}. Other authors suggest that there exist two relations for two different formation mechanisms \citep[gaseous formation mechanism and dry mergers,][]{2013ApJ...764..151G}, meaning that the replacement from the relation is not necessarily caused by undermassive black holes or overluminous bulges but by different formation processes. In particular, the $M_\mathrm{BH}-L_\mathrm{bulge}$ relation for galaxies which originate from gaseous formation mechanisms has a large scatter. Within this scatter, our galaxy might follow the relation. However, in our previous study \citep{2014A&A...561A.140B} with 11 objects, we show that LLQSOs (like HE 1029-1831) are systematically shifted from the Graham\&Scott-relation to the right.

We emphasize that both explanations are not inconsistent with our findings since in both scenarios the central component (``pseudo-bulge'' or ``S\'ersic spheroid'') is expected to be rich in gas and to contain younger stellar populations than classical (dry) bulges.

\begin{figure}
\includegraphics[width=\columnwidth]{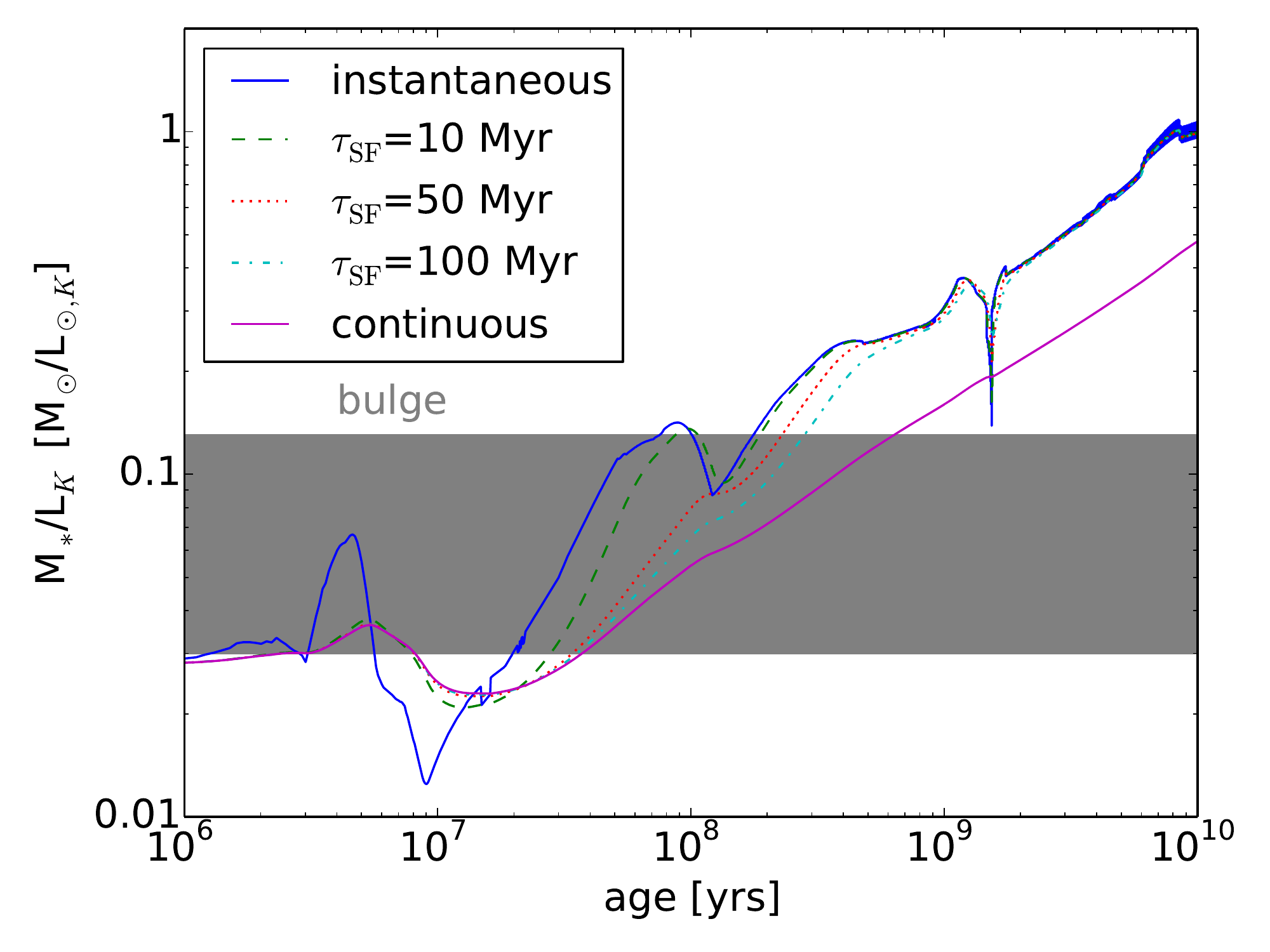}
\caption{$K$-band mass-to-light ratio as a function of the age of the stellar population, result from the \textsc{Starburst99} simulations. The $M_*/L_K$ ratio range derived from the dynamical mass and the bulge-disk decomposition (see Sect. \ref{sec:sf}) is indicated in gray.}
\label{fig:sb99_ml}
\end{figure}

\section{Summary and conclusions}
\label{sec:summary}

We have observed the inner $6\farcs4$ of the low-luminosity QSO HE 1029-1831 without AO and the inner $2\farcs4$ at a resolution of $0\farcs13=110\,\mathrm{pc}$ with AO, using the integral-field spectrograph SINFONI mounted at the VLT UT4 in Paranal. The main results of our analysis are the following:

\begin{itemize}
\item The \ion{H}{ii} emission shows a different flux distribution than the stellar continuum. Two gas spiral arms are located within the stellar bar. At smaller scales, a patchy circumnuclear ring is seen.

\item The flux distribution of the shock tracer [\ion{Fe}{ii}] follows the \ion{H}{ii} insofar as the emission is strongest in the points where the gas spiral arms meet the ring. Diagnostic line ratios as e.g. $\log(\mathrm{H}_2/\mathrm{Br}\gamma)$, $\log([\ion{Fe}{ii}]/\mathrm{Br}\gamma)$, or molecular hydrogen line ratios are indicative for star formation regions.

\item The mass of ionized hydrogen in the inner $3.2\,\mathrm{kpc}$ is $M_{\ion{H}{ii}}=2.9\times 10^7\,M_\odot$. The mass of cold molecular hydrogen within $1.2\,\mathrm{kpc}$ is $M_{\mathrm{H}_2\mathrm{,cold}}=(1.4-7.5)\times 10^9\,M_\odot$. This is in good agreement with H$_2$-mass estimates from CO measurements and comparable to gas masses derived for NUGA galaxies. We conclude that HE 1029-1831 has a massive reservoir of gas that is available for star formation and AGN fueling.

\item A bulge-disk decomposition with \textsc{Budda} reveals that the bulge luminosity does not follow published $M_\mathrm{BH}-L_\mathrm{bulge}$ relations for inactive galaxies. This is expected for LLQSOs \citep{2014A&A...561A.140B}. The dynamical mass of the bulge, however, is in good agreement with common $M_\mathrm{BH}-M_\mathrm{bulge}$ relations.

\item From the $M_*/L_K$ ratio and diagnostics like the Br$\gamma$ equivalent width and the supernova rates, we find evidence for the presence of young/intermediate-age stellar populations in the circumnuclear region. The favored model is an exponentially decaying starburst with time scale 50-100 Myr that began around 100-200 Myr ago. The light of the bulge component is dominated by this intermediate-age stellar population.
\end{itemize}

We conclude that circumnuclear star formation is a crucial factor in HE 1029-1831 that has to be taken into account for any thorough analysis \citep[see also, e.g.,][]{2004MNRAS.352..399J,2004MNRAS.355..273C}. We find evidence that, at least in the case of HE 1029-1831, the offset from the $M_\mathrm{BH}-L_\mathrm{bulge}$ relations could be explained by an overluminosity of the bulge component due to stellar populations that are younger than commonly observed in bulges.

As pointed out in Sect. \ref{sec:bh}, the Eddington ratio ($\lambda \gtrsim 0.06$) is  high, meaning that the black hole is in a growing phase. Though, since the black hole mass and - at least the dynamical - mass obey published $M_\mathrm{BH}-M_\mathrm{bulge}$ relations, from our data it seems rather unlikely that the offset from the $M_\mathrm{BH}-L_\mathrm{bulge}$ relation is caused by a significantly undermassive black hole that would grow ``towards'' the $M_\mathrm{BH}-L_\mathrm{bulge}$ relation.

With a larger sample of LLQSOs that have high-resolution data, in spatial and spectral dimension, we will be able to further investigate the star-formation properties in the circumnuclear regions.

\begin{acknowledgements}
The authors thank the anonymous referee for the helpful report. G.~Busch thanks Marcus Bremer for many fruitful discussions. This work was supported in part by the Deutsche Forschungsgemeinschaft (DFG) via SFB 956. G.~Busch is member of the Bonn-Cologne Graduate School of Physics and Astronomy (BCGS). S.~Smaji\'c is member of the International Max Planck Research School (IMPRS) for Astronomy and Astrophysics Bonn/Cologne. J.~Scharw\"achter acknowledges the European Research Council for the Advanced Grant Program Number 267399-Momentum. M.~Valencia-S.~received funding from the European Union Seventh Framework Programme (FP7/2007-2013) under grant agreement No. 312789.
\end{acknowledgements}

\bibliographystyle{aa}
\bibliography{he1029} 

\begin{appendix}
\section{Appendix}

\begin{figure*}
\includegraphics[width=0.45\linewidth]{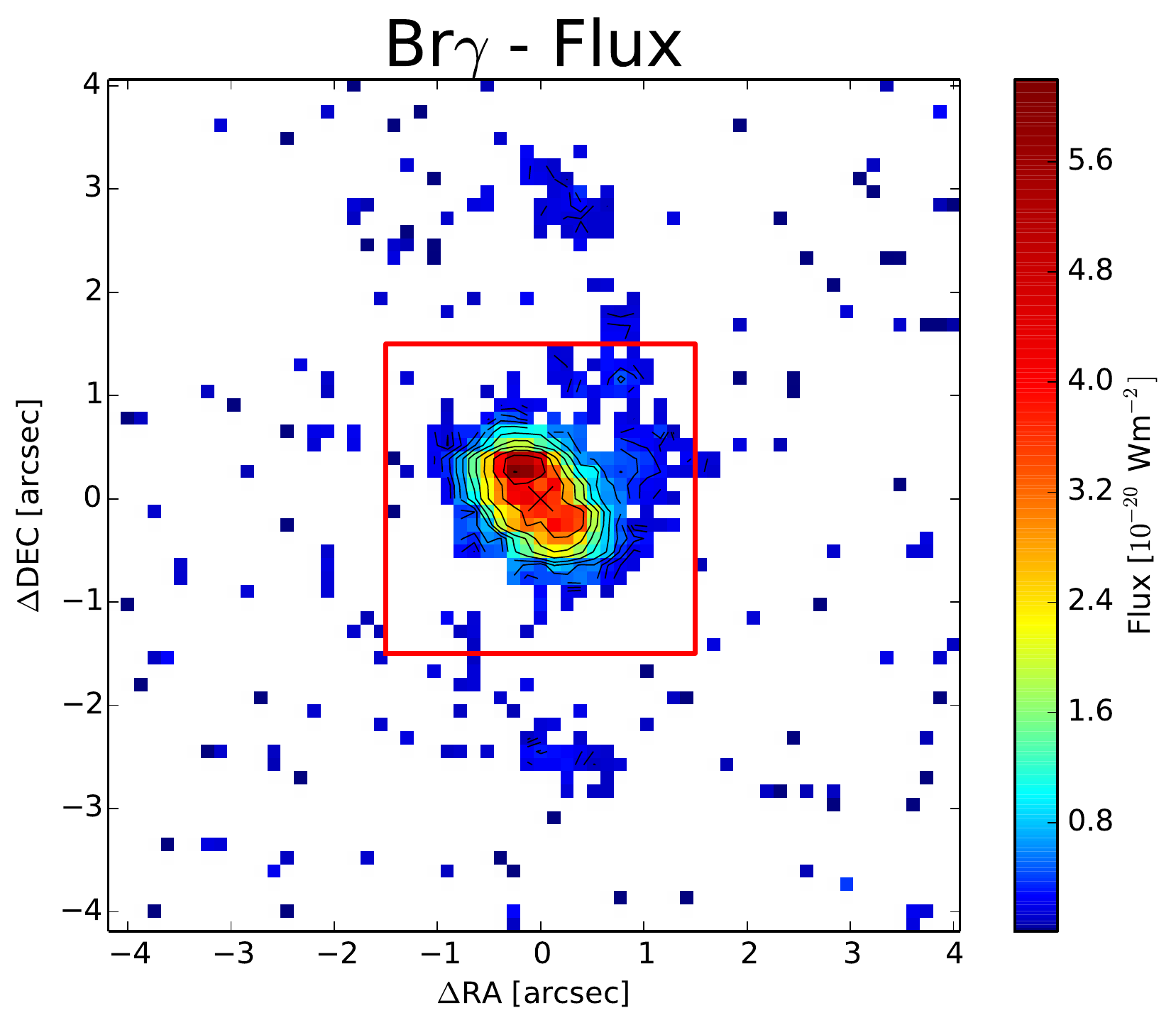}
\includegraphics[width=0.45\linewidth]{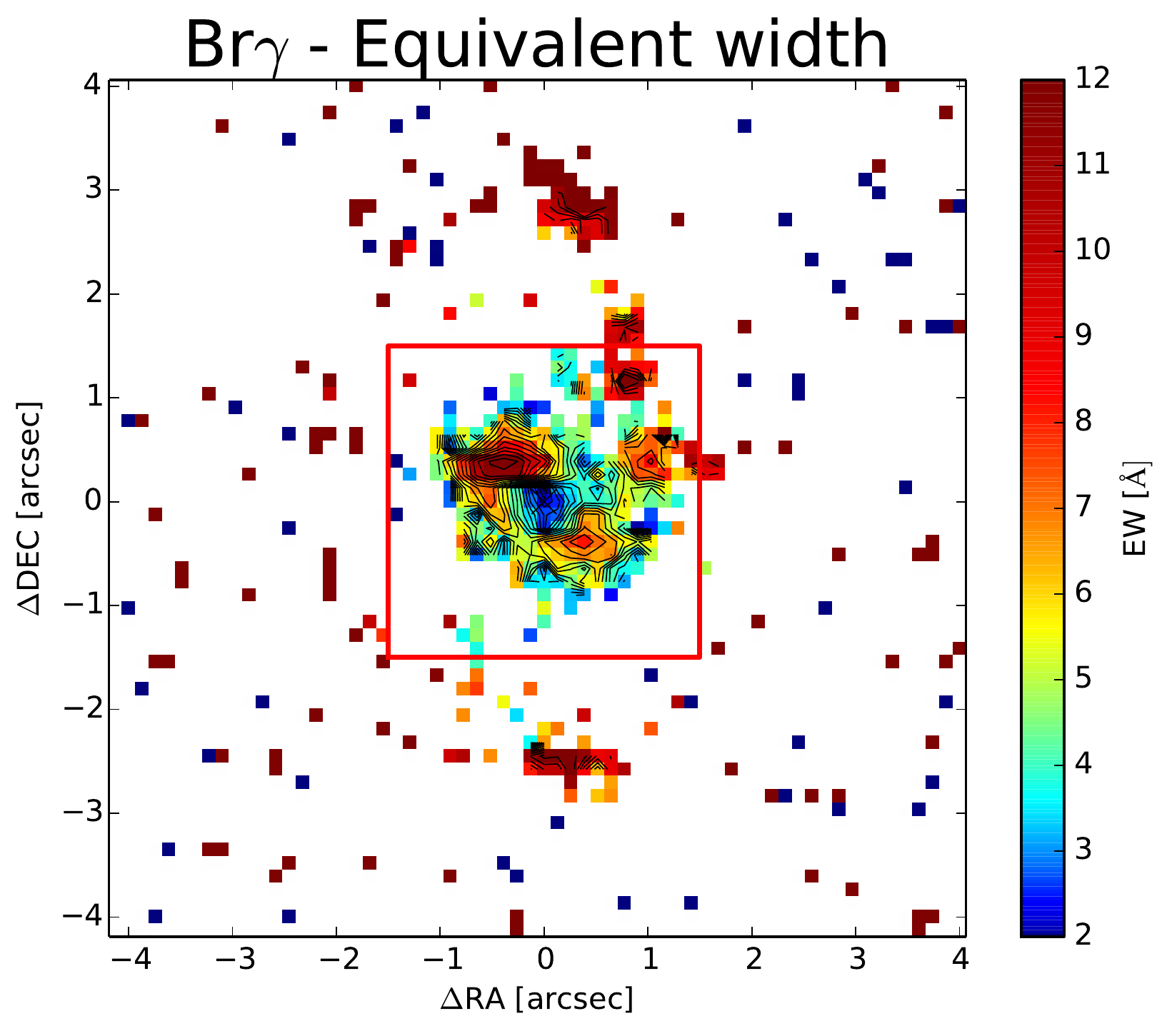}

\includegraphics[width=0.45\linewidth]{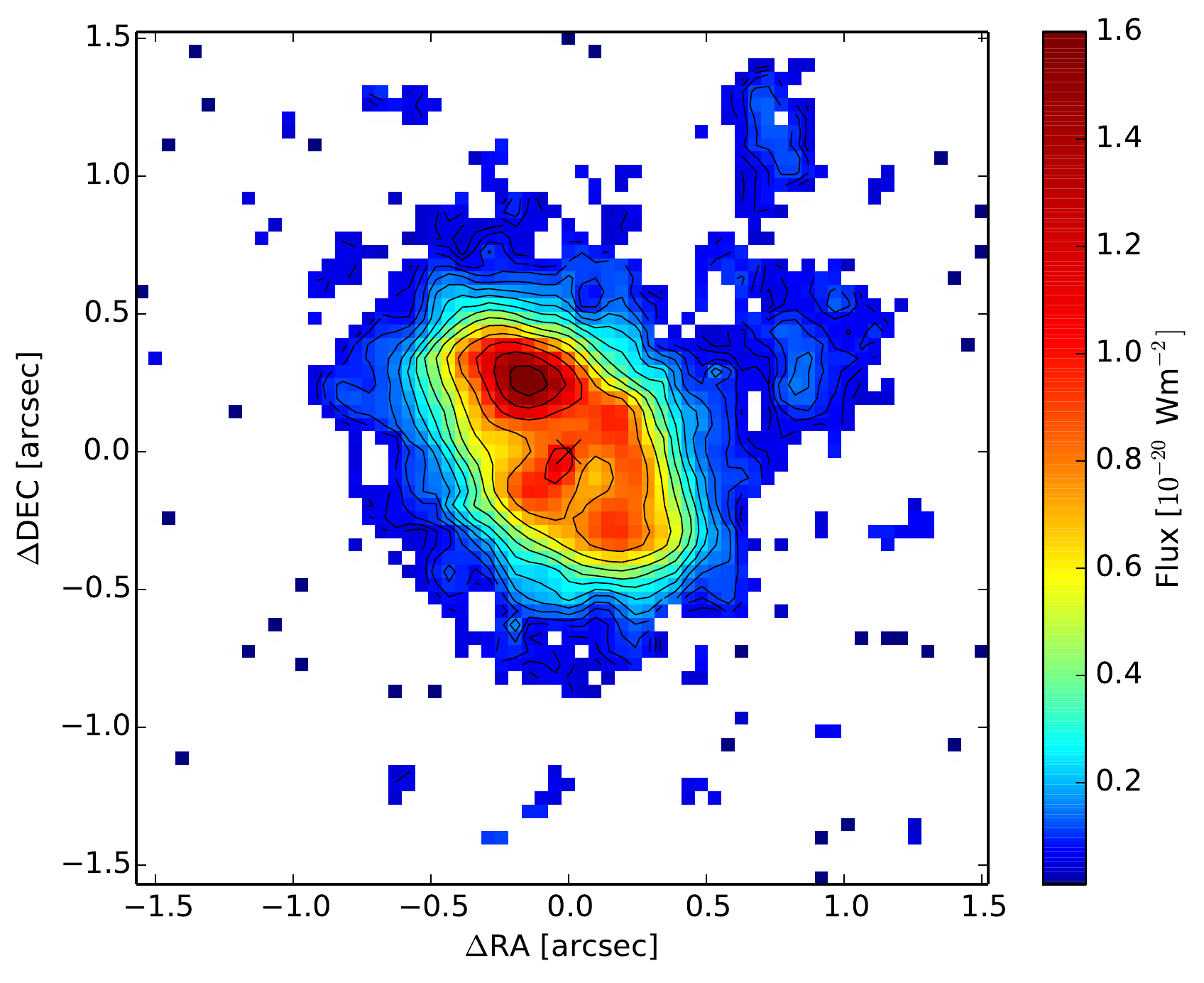}
\includegraphics[width=0.45\linewidth]{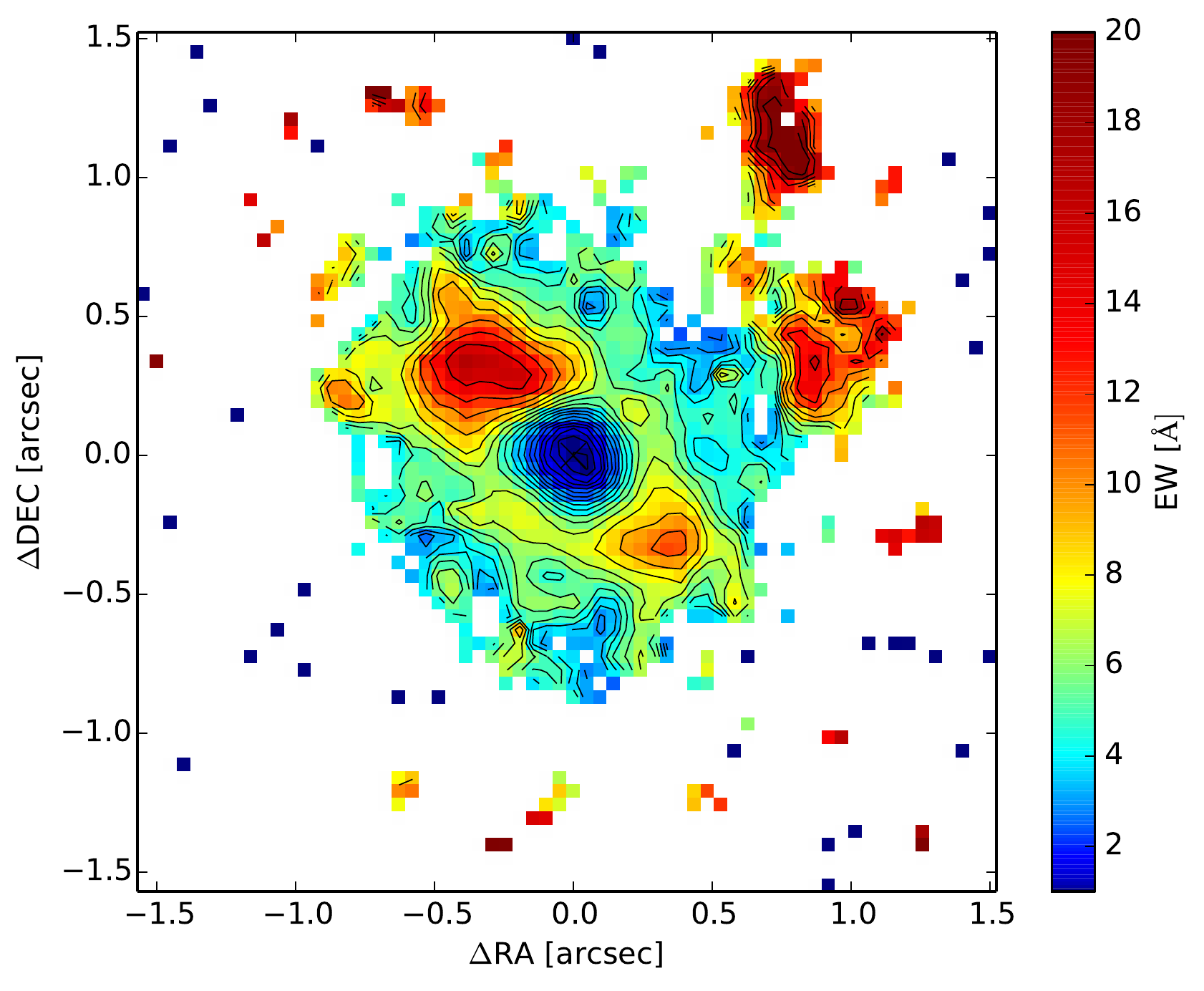}
\caption{The same as Fig. \ref{fig:paa} but for Br$\gamma$.}
\label{fig:brg}
\end{figure*}
\end{appendix}

\end{document}